\documentclass[twocolumn]{autart}

    \usepackage{jhoagg}
    \usepackage{amsmath}
    \usepackage{mathtools}
    \usepackage{amsfonts}
    \usepackage{amssymb,latexsym}
    \usepackage[psamsfonts]{eucal}
    \usepackage{graphicx}                               
    \usepackage[dvips]{epsfig}
    \usepackage{epstopdf}
    \usepackage[inline]{enumitem}
    \usepackage{psfrag}
    \usepackage{indentfirst}
    \usepackage{setspace}
    \usepackage{float}
    \usepackage{cite}
    \usepackage{accents}
    \usepackage{comment}
    \usepackage{xcolor}
    \usepackage[ruled, lined, norelsize]{algorithm2e}

 \makeatletter
 \g@addto@macro\normalsize{%
   \setlength\abovedisplayskip{4pt}
   \setlength\belowdisplayskip{4pt}
   \setlength\abovedisplayshortskip{4pt}
   \setlength\belowdisplayshortskip{4pt}
 }
 \makeatother
 \allowdisplaybreaks

    \graphicspath{ {Figures/} }
    \DeclareGraphicsExtensions{.png}

    \newtheorem{theorem}{\indent Theorem}

    \newtheorem{proposition}{\indent Proposition}

    \newtheorem{assumption}{\indent Assumption}

    \newtheorem{remark}{\indent Remark}

    \newtheorem{lemma}{\indent Lemma}

    \newtheorem{corollary}{\indent Corollary}

    \newtheorem{definition}{\indent Definition}

    \newtheorem{example}{\indent Example}

    \newcommand\xqed[1]{%
      \leavevmode\unskip\penalty9999 \hbox{}\nobreak\hfill
      \quad\hbox{#1}}
    \newcommand\exampletriangle{\xqed{$\triangle$}}

    \newcommand{\ubar}[1]{\underaccent{\bar}{#1}}

    \newlist{enumalph}{enumerate}{1}
    \setlist[enumalph]{label=\textit{(\alph*)}}
    
    \newlength\fheight
    \newlength\fwidth

    \usepackage{tikz}
    \usepackage{pgf}
    \usepackage{pgfplots}
    \usepgflibrary{plothandlers}
    \usepgfplotslibrary{external} 
    \pgfkeys{/pgf/number format/.cd,1000 sep={}}  
    \usetikzlibrary{calc,shapes.misc,plotmarks}
    \pgfplotsset{compat=1.13}
    
    \let\originalleft\left
    \let\originalright\right
    \renewcommand{\left}{\mathopen{}\mathclose\bgroup\originalleft}
    \renewcommand{\right}{\aftergroup\egroup\originalright}

\begin{document}

\begin{frontmatter}
\runtitle{Soft-Minimum and Soft-Maximum Barrier Functions}

\title{Soft-Minimum and Soft-Maximum Barrier Functions for\\ Safety with Actuation Constraints}

\thanks[footnoteinfo]{This work is supported in part by the National Science Foundation (1849213,1932105) and the Air Force Office of Scientific Research (FA9550-20-1-0028).}

\author{Pedram Rabiee}\ead{pedram.rabiee@uky.edu},
\author{Jesse B. Hoagg}\ead{jesse.hoagg@uky.edu}

\address{Department of Mechanical and Aerospace Engineering, University of Kentucky, Lexington, KY 40506}


\begin{abstract}                          
This paper presents two new control approaches for guaranteed safety (remaining in a safe set) subject to actuator constraints (the control is in a convex polytope). 
The control signals are computed using real-time optimization, including linear and quadratic programs subject to affine constraints, which are shown to be feasible.  
The first control method relies on a soft-minimum barrier function that is constructed using a finite-time-horizon prediction of the system trajectories under a known backup control. 
The main result shows that the control is continuous and satisfies the actuator constraints, and a subset of the safe set is forward invariant under the control. 
Next, we extend this method to allow from multiple backup controls. 
This second approach relies on a combined soft-maximum/soft-minimum barrier function, and it has properties similar to the first.
We demonstrate these controls on numerical simulations of an inverted pendulum and a nonholonomic ground robot.
\end{abstract}

\end{frontmatter}

\section{Introduction}
Robots and autonomous systems are often required to respect safety-critical constraints while achieving a specified task \cite{borrmann2015control,nguyen2015safety}. 
Safety constraints can be achieved by determining a control that makes a designated safe set $\SSS_\rms \subset \BBR^n$ forward invariant with respect to the closed-loop dynamics~\cite{blanchini1999set}, that is, designing a control for which the state is guaranteed to remain in $\SSS_\rms$.
Approaches that address safety using set invariance include reachability methods~\cite{chen2018hamilton,herbert2021scalable}, model predictive control~\cite{wabersich2022predictive,koller2018learning,zeng2021safety}, and barrier function (BF) methods (e.g.,~\cite{prajna2007framework,panagou2015distributed,tee2009barrier,jin2018adaptive,ames2014control,ames2016control,jankovic2018robust}).

Barrier functions are employed in a variety of ways.
For example, they are used for Lyapunov-like control design and analysis \cite{prajna2007framework,panagou2015distributed,tee2009barrier,jin2018adaptive}.
As another example, the control barrier function (CBF) approaches in~\cite{ames2014control,ames2016control,jankovic2018robust} compute the control signal using real-time optimization.
These optimization-based methods can be modular in that they often combine a nominal performance controller (which may not attempt to respect safety) with a safety filter that performs a real-time optimization using CBF constraints to generate a control that guarantees safety. 
This real-time optimization is often formulated as an instantaneous minimum-intervention problem, that is, the problem of finding a control at the current time instant that is as close as possible to the nominal performance control while satisfying the CBF safety constraints.

Barrier-function methods typically rely on the assumption that $\SSS_\rms$ is control forward invariant (i.e., there exists a control that makes $\SSS_\rms$ forward invariant). 
For systems without actuator constraints (i.e., input constraints), control forward invariance is satisfied under relatively minor structural assumptions (e.g., constant relative degree). 
In this case, the control can be generated from a quadratic program that employs feasible CBF constraints (e.g., \cite{ames2014control,ames2016control,jankovic2018robust}).
In contrast, actuator constraints can prevent $\SSS_\rms$ from being control forward invariant.
In this case, it may be possible to compute a control forward invariant subset of $\SSS_\rms$ using methods such as Minkowski operations~\cite{rakovic2004computation}, sum-of-squares~\cite{korda2014convex,xu2017correctness}, approximate solutions of a Hamilton-Jacobi partial differential equation~\cite{mitchell2005time}, or sampling~\cite{gillula2014sampling}. 
However, these methods may not scale to high-dimensional systems.

Another approach to address safety with actuator constraints is to use a prediction of the system trajectories into the future to obtain a control forward invariant subset of $\SSS_\rms$. 
For example,~\cite{squires2018constructive} uses the trajectory under a backup control.
However,~\cite{squires2018constructive} uses an infinite time horizon prediction, which limits applicability. 
In contrast,~\cite{gurrietScalableSafety2020,chen2020guaranteed} determine a control forward invariant subset of $\SSS_\rms$ from a BF constructed from a finite-horizon prediction under a backup control.
This BF uses the minimum function, which is not continuously differentiable and cannot be used directly to form a BF constraint for real-time optimization.
Thus,~\cite{gurrietScalableSafety2020,chen2020guaranteed} replace the original BF by a finite number of continuously differentiable BFs.
However, the number of substitute BFs (and thus optimization constraints) increases as the prediction horizon increases, and these multiple BF constraints can be conservative.
It is also worth noting that~\cite{gurrietScalableSafety2020,chen2020guaranteed} do not guarantee feasibility of the optimization with these multiple BF constraints.
Related approaches are in~\cite{xiao2022sufficient, singletary2022onboard,singletary2022safe}.

This paper makes several new contributions. First, we present a soft-minimum BF that uses a finite-horizon prediction of the system trajectory under a backup control.
We show that this BF describes a control forward invariant (subject to actuator constraints) subset of $\SSS_\rms$.
Since the soft-minimum BF is continuously differentiable, it can be used to form a single non-conservative BF constraint regardless of the prediction horizon. 
The soft-minimum BF facilitates the paper's second contribution, namely, a real-time optimization-based control that guarantees safety with actuator constraints.
Notably, the optimization required to compute the control is convex with guaranteed feasibility.
Next, we extend this approach to allow from multiple backup controls by using a novel soft-maximum/soft-minimum BF. 
In comparison to the soft-minimum BF, the soft-maximum/soft-minimum BF (with multiple backup controls) can yield a larger control forward invariant subset of $\SSS_\rms$. 
Some preliminary results on the soft-minimum BF appear in \cite{rabiee2023soft}.

\section{Notation}

Let $\rho>0$, and consider $\mbox{softmin}_\rho, \mbox{softmax}_\rho: \BBR^N \to \BBR$ defined by 
\begin{align*}
\mbox{softmin}_\rho (z_1,\ldots,z_N) &\triangleq -\frac{1}{\rho}\log\sum_{i=1}^Ne^{-\rho z_i},\\
\mbox{softmax}_\rho ( z_1,\ldots,z_N ) &\triangleq \frac{1}{\rho}\log\sum_{i=1}^Ne^{\rho z_i} - \frac{\log N }{\rho},
\end{align*}
which are the \textit{soft minimum} and \textit{soft minimum}.
The next result relates soft minimum and soft maximum to the minimum and maximum.

\begin{proposition} \label{fact:softmin_bound}
\rm{
Let $z_1,\ldots, z_N \in \BBR$. 
Then,
\begin{align*}
 \min \, \{ z_1,\ldots, z_N \} - \frac{\log N }{\rho} 
 &\leq \mbox{softmin}_\rho(z_1,\ldots,z_N) \\
 &\leq \min \, \{z_1,\ldots, z_N\},
\end{align*}
and
\begin{align*} 
 \max\,\{z_1,\ldots, z_N\}  - \frac{\log N}{\rho}
 &\leq \mbox{softmax}_\rho(z_1,\ldots,z_N) \\
 &\leq \max \, \{z_1,\ldots, z_N\}.
\end{align*}
}
\end{proposition}

Proposition~\ref{fact:softmin_bound} shows that as $\rho \to \infty$, $\mbox{softmin}_\rho$ and $\mbox{softmax}_\rho$ converge to the minimum and maximum. 
Thus, $\mbox{softmin}_\rho$ and $\mbox{softmax}_\rho$ are smooth approximations of the minimum and maximum.
Note that if $N>1$, then the soft minimum is strictly less than the minimum.

For a continuously differentiable function $\eta \colon \BBR^n \to \BBR^l$, let $\eta^\prime \colon \BBR^n \to \BBR^{l \times n}$ be defined by $\eta^\prime(x) = \frac{\partial \eta(x)}{\partial x}$. 
The Lie derivatives of $\eta$ along the vector field of $\psi \colon \BBR^n \to \BBR^{n \times p}$ is $L_\psi \eta(x) \triangleq \eta^\prime(x) \psi(x)$. 
Let $\mbox{int }\SA$, $\mbox{bd }\SA$, $\mbox{cl }\SA$ denote the interior, boundary, and closure of the set $\SA \subset \BBR^n$.
Let $a,b \in \BBR^{r}$. If each element of $a$ is less than or equal to the corresponding element of $b$, then we write $a \preceq b$.

\section{Problem Formulation}

Consider 
\begin{equation}\label{eq:dynamics}
\dot x(t) = f(x(t))+g(x(t)) u(t),
\end{equation}
where $f:\BBR^n \to \BBR^n$ and $g:\BBR^n\to \BBR^{n\times m}$ are continuously differentiable on $\BBR^n$, $x(t) \in \BBR^{n}$ is the state, $x(0) = x_0 \in \BBR^n$ is the initial condition, and $u(t) \in \BBR^m$ is the control.
Let $A_u \in \BBR^{r \times m}$ and $b_u \in \BBR^{r}$, and define
\begin{equation}
    \SU \triangleq \{u \in \BBR^m: A_u u \preceq b_u\} \subset \BBR^{m},
    \label{eq:SU}
\end{equation}
which we assume is bounded and not empty.
We call $u$ an \textit{admissible control} if for all $t \ge 0$, $u(t) \in \SU$.

Let $h_\rms:\BBR^n \to \BBR$ be continuously differentiable, and define the \textit{safe set}
\begin{equation}\label{eq:S_s}
\SSS_\rms \triangleq \{x \in \BBR^n \colon h_\rms(x) \geq 0 \}.
\end{equation}
Note that $\SSS_\rms$ is not assumed to be control forward invariant with respect to \eqref{eq:dynamics} where $u$ is an admissible control.
In other words, there may not exist an admissible control $u$ such that if $x_0 \in \SSS_\rms$, then for all $t \ge 0$, $x(t) \in \SSS_\rms$.

Next, consider the \textit{nominal desired control} $u_\rmd : \BBR^n \to \BBR^m$ designed to satisfy performance specifications, which can be independent of and potentially conflict with safety.
Thus, $\SSS_\rms$ is not necessarily forward invariant with respect to \eqref{eq:dynamics} where $u = u_\rmd$. 
We also note that $u_\rmd$ is not necessarily an admissible control.

The objective is to design a full-state feedback control $u:\BBR^n \to \BBR^m$ such that for all initial conditions in a subset of $\SSS_\rms$, the following hold:

\begin{enumerate}[leftmargin=1cm]
	\renewcommand{\labelenumi}{(O\arabic{enumi})}
	\renewcommand{\theenumi}{(O\arabic{enumi})}

\item\label{obj1}
For all $t \ge 0$, $x(t) \in \SSS_\rms$.

\item\label{obj2}
For all $t \ge 0$, $u(x(t)) \in \SU$.

\item\label{obj3}
For all $t \ge 0$, $\| u(x(t)) -u_\rmd(x(t)) \|_2$ is small.
\end{enumerate}

\section{Barrier Functions Using the Trajectory Under a Backup Control}\label{section:softmin_prelim}

Consider a continuously differentiable \textit{backup control} $u_\rmb : \BBR^n \to \SU$.
Let $h_\rmb : \BBR^n \to \BBR$ be continuously differentiable, and 
define the \textit{backup safe set}
\begin{equation}\label{eq:S_b}
\SSS_\rmb \triangleq \{x \in \BBR^n \colon h_\rmb(x) \geq 0 \}.
\end{equation}
We assume $\SSS_\rmb \subseteq \SSS_\rms$ and make the following assumption.

\begin{assumption}\label{assump:S_b_contract}\rm
If $u=u_\rmb$ and $x_0 \in \SSS_\rmb$, then for all $t \ge 0$, $x(t) \in \SSS_\rmb$. 
\end{assumption}

Assumption~\ref{assump:S_b_contract} states that $\SSS_\rmb$ is forward invariant with respect to \eqref{eq:dynamics} where $u = u_\rmb$.
However, $\SSS_\rmb$ may be small relative to $\SSS_\rms$.

Consider $\tilde f: \BBR^n \to \BBR^n$ defined by 
\begin{equation}\label{eq:ftilde}
\tilde f(x) \triangleq f(x) + g(x) u_\rmb(x),
\end{equation}
which is the right-hand side of the closed-loop dynamics \eqref{eq:dynamics} with $u=u_\rmb$. 
Next, let $\phi: \BBR^n \times [0,\infty) \to \BBR^n$ satisfy 
\begin{equation}\label{eq:phi_def}
    \phi(x, \tau) = x +\int_{0}^{\tau} \tilde f(\phi(x,\sigma)) \, \rmd \sigma,
\end{equation}
which implies that $\phi(x,\tau)$ is the solution to \eqref{eq:dynamics} at time $\tau$ with $u=u_\rmb$ and initial condition $x$.

Let $T > 0$ be a time horizon, and consider $h_*:\BBR^n \to \BBR$ defined by
\begin{equation}
h_*(x) \triangleq \min \, \left \{h_\rmb(\phi(x,T)), \min_{\tau \in [0,T]} h_\rms(\phi(x,\tau)) \right \}, \label{eq:h_min_cont_def}
\end{equation}
and define
\begin{equation}
\SSS_* \triangleq \{ x \in \BBR^n \colon h_*(x) \ge 0 \}\label{eq:defS_cont}.
\end{equation} 
For all $x \in \SSS_*$, the solution \eqref{eq:phi_def} under $u_\rmb$ does not leave $\SSS_\rms$ and reaches $\SSS_\rmb$ by time $T$.
The next result relates $\SSS_*$ to $\SSS_\rmb$ and $\SSS_\rms$. The result is similar to \cite[Proposition 6]{gurrietScalableSafety2020}.

\begin{proposition} \label{prop:S*}\rm
Assume that $u_\rmb$ satisfies Assumption~\ref{assump:S_b_contract}.
Then, $\SSS_\rmb \subseteq \SSS_* \subseteq \SSS_\rms$.
\end{proposition}

\begin{pf}
Let $x_1 \in \SSS_\rmb$.
Assumption~\ref{assump:S_b_contract} implies for all $t \ge 0$, $\phi(x_1,t) \in \SSS_\rmb \subseteq \SSS_\rms$, which implies for all $t \ge 0$, $h_\rms(\phi(x_1,t)) \ge 0$ and  $h_\rmb(\phi(x_1,t)) \ge 0$. 
Thus, it follows from~\eqref{eq:h_min_cont_def} and~\eqref{eq:defS_cont} that $h_*(x_1) \ge 0$, which implies $x_1 \in \SSS_*$. 
Therefore, $\SSS_\rmb \subseteq \SSS_*$.

Let $x_2 \in \SSS_*$, and \eqref{eq:h_min_cont_def} implies $h_\rms(x_2) = h_\rms(\phi(x_2,0)) \ge h_*(x_2) \ge 0$. 
Thus, $x_2 \in \SSS_\rms$, which implies $\SSS_* \subseteq \SSS_\rms$. 
{\hfill$\Box$}\end{pf}

The next result shows that $\SSS_*$ is forward invariant with respect to \eqref{eq:dynamics} where $u=u_\rmb$. 
In fact, this result shows that the state converges to $\SSS_\rmb \subseteq \SSS_*$ by time $T$. 

\begin{proposition} \label{prop:fwd_invar_Sc}
\rm 
Consider \eqref{eq:dynamics}, where $x_0 \in \SSS_*$, $u = u_\rmb$, and $u_\rmb$ satisfies Assumption~\ref{assump:S_b_contract}.
Then, the following hold:
\begin{enumalph}
\item For all $t \ge T$, $x(t) \in \SSS_\rmb$. \label{fact:fwd_invar_Sc_Sb}
\item For all $t \ge 0$, $x(t) \in \SSS_*$. \label{fact:fwd_invar_Sc_traj}
\end{enumalph}
\end{proposition}

\begin{pf}
To prove~\ref{fact:fwd_invar_Sc_Sb}, since $x_0 \in \SSS_*$, it follows from~\eqref{eq:h_min_cont_def} and~\eqref{eq:defS_cont} that $h_\rmb(\phi(x_0, T)) \ge 0$, which implies $x(T) = \phi(x_0, T) \in \SSS_\rmb$. 
Since $x(T) \in \SSS_\rmb$, Assumption~\ref{assump:S_b_contract} implies for all $t \ge T$, $x(t) \in \SSS_\rmb$, which confirms \ref{fact:fwd_invar_Sc_Sb}.

To prove~\ref{fact:fwd_invar_Sc_traj}, let $t_1 \ge 0$ and consider 2 cases: $t_1 \ge T$, and $t_1 < T$.
First, let $t_1 \ge T$, and it follows from \ref{fact:fwd_invar_Sc_Sb} that for all $t \ge t_1$, $x(t) \in \SSS_\rmb \subseteq \SSS_\rms$.
Since, in addition, for all $t \ge t_1$, $x(t) = \phi(x(t_1),t-t_1)$, it follows from~\eqref{eq:h_min_cont_def} and~\eqref{eq:defS_cont} that $h_*(x(t_1)) \ge 0$, which implies $x(t_1) \in \SSS_*$. 
Next, let $t_1 < T$.
Since $x_0 \in \SSS_*$, it follows from~\eqref{eq:h_min_cont_def} and~\eqref{eq:defS_cont} that for all $t\in[t_1,T]$, $h_\rms(\phi(x_0, t)) \ge 0$, which implies for all $t\in[t_1,T]$, $x(t) = \phi(x_0, t) \in \SSS_\rms$.
Since, in addition, for all $t \ge T$, $x(t) \in \SSS_\rmb \subseteq \SSS_\rms$, it follows from from~\eqref{eq:h_min_cont_def} and~\eqref{eq:defS_cont} that $h_*(\phi(x_0, t_1)) \ge 0$, which implies $x(t_1) \in \SSS_*$. 
{\hfill$\Box$}\end{pf}

Proposition~\ref{prop:fwd_invar_Sc} implies that for all $x_0 \in \SSS_*$, the backup control $u_\rmb$ satisfies \ref{obj1} and \ref{obj2}. 
However, $u_\rmb$ does not address \ref{obj3}.
One approach to address \ref{obj3} is to use $h_*$ as a BF in a minimum intervention quadratic program. 
However, $h_*$ is not continuously differentiable.
Thus, it cannot be used directly to construct a BF constraint because the constraint and associated control would not be well-defined at the locations in the state space where $h_*$ is not differentiable. 
This issue is addressed in \cite{gurrietScalableSafety2020} by using multiple BFs---one for each argument of the minimum in \eqref{eq:h_min_cont_def}.
However, \eqref{eq:h_min_cont_def} has infinitely many arguments because the minimum is over $[0,T]$.
Thus, \cite{gurrietScalableSafety2020} uses a sampling of times. 
Specifically, let $N$ be a positive integer, and define $\SN \triangleq \{ 0, 1, \ldots, N \}$ and $T_\rms \triangleq T/N$. 
Then, consider $\bar h_*:\BBR^n \to \BBR$ defined by
\begin{equation}
\bar h_*(x) \triangleq \min \, \left \{h_\rmb(\phi(x,NT_\rms)), \min_{i \in \SN } h_\rms(\phi(x,iT_\rms)) \right \},
\label{eq:h_min_def}
\end{equation}
and define
\begin{equation}
\bar \SSS_* \triangleq \{ x \in \BBR^n \colon \bar h_*(x) \ge 0 \}. \label{eq:defS_star}
\end{equation}
The next result relates $\bar \SSS_*$ to $\SSS_*$ and $\SSS_\rms$.

\begin{proposition} \label{prop:bar S*}
\rm $\SSS_* \subseteq \bar \SSS_* \subseteq \SSS_\rms$.
\end{proposition}

\begin{pf}
Let $x_1 \in \SSS_*$, and it follows from
from~\eqref{eq:h_min_cont_def}--\eqref{eq:defS_star} that $x_1 \in \bar \SSS_*$, which implies $\SSS_* \subseteq \bar \SSS_*$.

Let $x_2 \in \bar \SSS_*$, and \eqref{eq:h_min_def} implies $h_\rms(x_2) = h_\rms(\phi(x_2,0)) \ge \bar h_*(x_2) \ge 0$. 
Thus, $x_2 \in \SSS_\rms$, which implies $\bar \SSS_* \subseteq \SSS_\rms$. 
{\hfill$\Box$}\end{pf}

The next result shows that for all $x_0 \in \bar \SSS_*$, the backup control $u_\rmb$ causes the state to remain in $\bar \SSS_*$ at the sample times $T_\rms,2T_\rms,\ldots,N T_\rms$ and converge to $\SSS_\rmb$ by time $T$.

\begin{proposition} \label{prop:fwd_invar_S*}
\rm 
Consider \eqref{eq:dynamics}, where $x_0 \in \SSS_*$, $u = u_\rmb$, and $u_\rmb$ satisfies Assumption~\ref{assump:S_b_contract}.
Then, the following hold:
\begin{enumalph}
\item For all $t \ge T$, $x(t) \in \SSS_\rmb$. \label{fact:fwd_invar_S*_Sb}
\item For all $i\in \SN$, $x(iT_\rms) \in \bar \SSS_*$. \label{fact:fwd_invar_S*_traj}
\end{enumalph}
\end{proposition}

\begin{pf}
To prove~\ref{fact:fwd_invar_S*_Sb}, since $x_0 \in \bar \SSS_*$, it follows from~\eqref{eq:h_min_def} and~\eqref{eq:defS_star} that $h_\rmb(\phi(x_0, T)) \ge 0$, which implies $x(T) = \phi(x_0, T) \in \SSS_\rmb$. 
Since $x(T) \in \SSS_\rmb$, Assumption~\ref{assump:S_b_contract} implies for all $t \ge T$, $x(t) \in \SSS_\rmb$, which confirms \ref{fact:fwd_invar_S*_Sb}.

To prove~\ref{fact:fwd_invar_Sc_traj}, let $i_1 \in \SN$. 
Since $x_0 \in \bar \SSS_*$, it follows from~\eqref{eq:h_min_def} and~\eqref{eq:defS_star} that for all $i\in\{i_1,\ldots,N\}$, $h_\rms(\phi(x_0, i T_\rms)) \ge 0$, which implies for all $i\in\{i_1,\ldots,N\}$, $x(i T_\rms) = \phi(x_0, i T_\rms) \in \SSS_\rms$.
Since, in addition, for all $t \ge N T_\rms$, $x(t) \in \SSS_\rmb \subseteq \SSS_\rms$, it follows from~\eqref{eq:h_min_def} and~\eqref{eq:defS_star} that 
$\bar h_*(\phi(x_0, i_1 T_\rms)) \ge 0$, which implies $x(i_1 T_\rms) \in \bar \SSS_*$. 
{\hfill$\Box$}\end{pf}

Proposition~\ref{prop:fwd_invar_S*} does not provide any information about the state in between the sample times. 
Thus, Proposition~\ref{prop:fwd_invar_S*} does not imply that $\bar \SSS_*$ is forward invariant with respect to \eqref{eq:dynamics} where $u=u_\rmb$. 
However, we can adopt an approach similar to \cite{gurrietScalableSafety2020} to determine a superlevel set of $\bar h_*$ such that for all initial conditions in that superlevel set, $u_\rmb$ keeps the state in $\SSS_*$ for all time. 
To define this superlevel set, let $l_\rms$ be the Lipschitz constant of $h_\rms$ with respect to the two norm, and define $l_\phi \triangleq \sup_{x\in \bar \SSS_*} \| \tilde f(x) \|_2$, which is finite if $\SSS_\rms$ is bounded. 
Define the superlevel set
\begin{equation}\label{eq:ubarS_*_def}
\ubar \SSS_* \triangleq \left \{x\in \BBR^n : \bar h_*(x) \ge \tfrac{1}{2} T_\rms l_\phi l_\rms \right \}.
\end{equation}
The next result combines \cite[Thm. 1]{gurrietScalableSafety2020} and Proposition~\ref{prop:bar S*}.

\begin{proposition}\label{prop:set_rel_softmin}
\rm $\ubar \SSS_* \subseteq \SSS_* \subseteq \bar \SSS_* \subseteq\SSS_\rms$. 
\end{proposition}

Together, Propositions~\ref{prop:fwd_invar_Sc} and \ref{prop:set_rel_softmin} imply that for all $x_0 \in \ubar \SSS_*$, the backup control $u_\rmb$ keeps the state in $\SSS_*$ for all time. 
However, $u_\rmb$ does not address \ref{obj3}.

Since $\bar h_*$ is not continuously differentiable, \cite{gurrietScalableSafety2020} addresses \ref{obj3} using a minimum intervention quadratic program with $N+1$ BFs---one for each of the arguments in \eqref{eq:h_min_def}. 
However, this approach has 3 drawbacks. 
First, the number of BFs increases as the time horizon $T$ increases or the sample time $T_\rms$ decreases (i.e., as $N$ increases). 
Thus, the number of affine constraints and computational complexity increases as $N$ increases. 
Second, although imposing an affine constraint for each of the $N+1$ BFs is sufficient to ensure that $\bar h_*$ remains positive, it is not necessary. 
These $N+1$ affine constraints are conservative and can limit the set of feasible solutions for the control. 
Third, \cite{gurrietScalableSafety2020} does not guarantee feasibility of the optimization used to obtain the control.

The next section uses a soft-minimum BF to approximate $\bar h_*$ and presents a control synthesis approach with guaranteed feasibility and where the number of affine constraints is fixed (i.e., independent of $N$).

\section{Safety-Critical Control Using Soft-Minimum Barrier Function with One Backup Control}\label{section:softmin}

This section presents a continuous control that guarantees safety subject to the constraint that the control is admissible (i.e., in $\SU$). 
The control is computed using a minimum intervention quadratic program with a soft-minimum BF constraint. 
The control also relies on a linear program to provide a feasibility metric, that is, a measure of how close the quadratic program is to becoming infeasible. 
Then, the control continuously transitions to the backup control $u_\rmb$ if the feasibility metric or the soft-minimum BF are less than user-defined thresholds.

Let $\rho_1 > 0$, and consider $h:\BBR^n \to \BBR$ defined by
\begin{align}
h(x) &\triangleq  \mbox{softmin}_{\rho_1} ( h_\rms(\phi(x,0)), h_\rms(\phi(x,T_\rms)), \ldots,\nn\\
&\qquad h_\rms(\phi(x,N T_\rms)), h_\rmb(\phi(x,NT_\rms))),\label{eq:h_softmin_def}
\end{align}
which is continuously differentiable. 
Define
\begin{equation}\label{eq:defS_softmin}
\SSS \triangleq \{ x \in \BBR^n \colon h(x) \ge 0 \}.
\end{equation}
Proposition~\ref{fact:softmin_bound} implies that for all $x \in \BBR^n$, $h(x) < \bar h_*(x)$. 
Thus, $\SSS \subset \bar \SSS_*$.
Proposition~\ref{fact:softmin_bound} also implies that for sufficiently large $\rho_1>0$, $h(x)$ is arbitrarily close to $\bar h_*(x)$. 
Thus, $h$ is a smooth approximation of $\bar h_*$. 
However, if $\rho_1 >0$ is large, then $\| h^\prime(x) \|_2$ is large at points where $\bar h_*$ is not differentiable.
Thus, selecting $\rho_1$ is a trade-off between the conservativeness of $h$ and the size of $\| h^\prime(x) \|_2$.

Next, let $\alpha >0$ and $\epsilon \in [0, \sup_{x\in\SSS} h(x))$. Consider  $\beta\colon \BBR^n \to \BBR$ defined by
\begin{equation}
\label{eq:feas_check}
\beta(x) \triangleq  L_f h(x) + \alpha (h(x)- \epsilon) + \max_{\hat u \in \SU} 
 L_g h(x)  \hat u,
\end{equation}
where $\beta$ exists because $\SU$ is not empty.
Define
\begin{equation}
\label{eq:feasible_set_def}
    \SB \triangleq \{ x \in\BBR^n \colon \beta(x) \ge 0 \},
\end{equation}
The next result follows immediately from~\eqref{eq:feas_check} and~\eqref{eq:feasible_set_def}.

\begin{proposition}\label{prop:feas_softmin}
{\rm
For all $x \in \SB$, there exists $\hat u \in \SU$ such that $L_f h(x) + L_g h(x) \hat u + \alpha (h(x) - \epsilon) \ge 0$.
}
\end{proposition}

Let $\kappa_h, \kappa_\beta > 0$ and consider $\gamma \colon \BBR^n \to \BBR$ defined by
\begin{equation}\label{eq:gamma_def}
\gamma(x) \triangleq \min \left\{\frac{h(x) - \epsilon}{\kappa_h}, \frac{\beta(x)}{\kappa_\beta} \right\},
\end{equation}
and define
\begin{equation}\label{eq:Gamma_def}
    \Gamma \triangleq \{ x \in\BBR^n \colon \gamma(x) \ge 0 \}.
\end{equation}
Note that $\Gamma \subseteq \SB$.
For all $x \in \Gamma$, define
\begin{subequations}\label{eq:qp_softmin}
\begin{align}
& u_*(x) \triangleq \underset{\hat u \in \SU}{\mbox{argmin}}  \, 
\|\hat u - u_\rmd(x)\|_2^2 \label{eq:qp_softmin.a}\\
& \text{subject to}\nn\\
& L_f h(x) + L_g h(x) \hat u + \alpha (h(x) - \epsilon) \ge 0. \label{eq:qp_softmin.b}
\end{align}
\end{subequations}
Since $\Gamma \subseteq \SB$, Proposition~\ref{prop:feas_softmin} implies that for all $x \in \Gamma$, the quadratic program \eqref{eq:qp_softmin} has a solution.

Consider a continuous function $\sigma:\BBR \to [0,1]$ such that for all $a \in (-\infty,0]$, $\sigma(a) =0$; for all $a \in [1,\infty)$, $\sigma(a) =1$; and $\sigma$ is strictly increasing on $a \in [0,1]$. 
The following example provides one possible choice for $\sigma$. 

\begin{example}\label{ex:sigma}\rm
Consider $\sigma:\BBR \to [0,1]$ given by
\begin{equation}
    \sigma(a) = \begin{cases}
    0, & \mbox{if } a \le 0, \\
    a, & \mbox{if } 0 < a < 1, \\
    1, & \mbox{if } a \ge 1. \\
    \end{cases}
    \tag*{\exampletriangle}   
\end{equation}
\end{example}

Finally, define the control 
\begin{equation}\label{eq:u_cases_softmin}
    u(x) = \begin{cases} [1-\sigma(\gamma(x))] u_\rmb(x) + \sigma(\gamma(x)) u_*(x), & \mbox{if } x \in \Gamma, \\ u_\rmb(x), & \mbox{else}.
    \end{cases}
\end{equation}
Since $h$ is continuously differentiable, the quadratic program \eqref{eq:qp_softmin} requires only the single affine constraint \eqref{eq:qp_softmin.b} as opposed to the $N+1$ constraints used in \cite{gurrietScalableSafety2020}.
Since \eqref{eq:qp_softmin} has only one affine constraint, we can define the feasible set $\SB$ as the zero-superlevel set of $\beta$, which is the solution to the linear program \eqref{eq:feas_check}. 
Since there is only one affine constraint, we can use the homotopy in \eqref{eq:u_cases_softmin} to continuously transition from $u_*$ to $u_\rmb$ as $x$ leaves $\Gamma$.

\begin{remark}\rm 
The control~\eqref{eq:h_softmin_def}--\eqref{eq:u_cases_softmin} is designed for the case where the relative degree of \eqref{eq:dynamics} and \eqref{eq:h_softmin_def} is one (i.e., $L_gh(x) \ne 0$). 
However, this control can be applied independent of the relative degree. 
If $L_gh(x) = 0$, then it follows from \eqref{eq:feas_check}--\eqref{eq:qp_softmin} that for all $x \in \Gamma$, the solution to the quadratic program \eqref{eq:qp_softmin} is the unconstrained minimizer, which is the desired control if $u_\rmd(x) \in \SU$.
In this case, \eqref{eq:u_cases_softmin} implies that $u$ is determined from a continuous blending of $u_\rmd$ and $u_\rmb$ based on $\gamma$ (i.e., feasibility of \eqref{eq:qp_softmin} and safety). 
We also note that the control~\eqref{eq:h_softmin_def}--\eqref{eq:u_cases_softmin} can be generalized to address the case where the relative degree exceeds one. 
In this case, the linear program \eqref{eq:feas_check} for feasibility and the quadratic program constraint \eqref{eq:qp_softmin.b} are replaced by the appropriate higher-relative-degree Lie derivative expressions (see \cite{rabiee2024closed,rabiee2023composition,xiao2021high}).
\end{remark}

The next theorem is the main result on the control~\eqref{eq:h_softmin_def}--\eqref{eq:u_cases_softmin} that uses the soft-minimum BF approach.

\begin{theorem}
\label{thm:softmin}
\rm 
Consider \eqref{eq:dynamics} and $u$ given by~\eqref{eq:h_softmin_def}--\eqref{eq:u_cases_softmin}, where $\SU$ given by \eqref{eq:SU} is bounded and nonempty, and $u_\rmb$ satisfies Assumption~\ref{assump:S_b_contract}.
Then, the following hold:
\begin{enumalph}

\item
$u$ is continuous on $\BBR^n$.
\label{thm:softmin_u_continuity}

\item
For all $x \in \BBR^n$, $u(x) \in \SU$. 
\label{thm:softmin_u_cond}

\item
Let $x_0 \in \bar \SSS_*$. 
Assume there exists $t_1 \ge 0$ such that $x(t_1) \in \rm{bd \,}\bar \SSS_*$. 
Then, there exists $\tau \in (0, T_\rms]$ such that $x(t_1+\tau) \in \bar \SSS_* \subseteq \SSS_\rms$.
\label{thm:softmin_return_to_S_star}

\item 
Let $\epsilon \ge \frac{1}{2} T_\rms l_\phi l_\rms$ and $x_0 \in \SSS_*$.
Then, for all $t \ge 0$, $x(t) \in \SSS_* \subseteq \SSS_\rms$. \label{thm:softmin_forward_inv}

\end{enumalph}
\end{theorem}

\begin{pf}
To prove~\ref{thm:softmin_u_continuity}, we first show that $u_*$ is continuous on $\Gamma$.
Define $J(x,\hat u) \triangleq \| \hat u - u_\rmd(x) \|_2^2$ and $\Omega(x) \triangleq \{\hat u \in \SU : L_f h(x) + L_g h(x) \hat u + \alpha (h(x) - \epsilon) \ge 0\}$.
Let $a\in\Gamma \subseteq \SB$, and Proposition~\ref{prop:feas_softmin} implies $\Omega(a)$ is not empty. 
Since, in addition,  $J(a,\hat u)$ is strictly convex, $\SU$ is convex, and \eqref{eq:qp_softmin.b} with $x=a$ is convex, it follows that $u_*(a)$ is unique. 
Next, since for all $x \in \Gamma$, $u_*(x) \subseteq \SU$ is bounded, it follows that $u_*(\Gamma)$ is bounded. 
Thus, $\mbox{cl } u_*(\Gamma)$ is compact.
Next, since $\SU$ is a convex polytope and \eqref{eq:qp_softmin.b} is affine in $\hat u$, it follows from \cite[Remark 5.5]{borrelliPredictiveControl2017} that $\Omega$ is continuous at $a$.
Finally, since $u_*(a)$ exists and is unique, $\mbox{cl } u_*(\Gamma)$ is compact, $\Omega$ is continuous at $a$, and $J$ is continuous on $a \times \Omega(a)$, it follows from \cite[Corollary 8.1]{hogan1973point} that $u_*$ is continuous at $a$.
Thus, $u_*$ is continuous on $\Gamma$.

Define $J_2(x,\hat u) \triangleq L_f h(x) + L_g h(x) \hat u + \alpha (h(x) - \epsilon)$, and note that \eqref{eq:feas_check} implies $\beta(x) = \max_{\hat u \in \SU} J_2(x,\hat u)$. 
Since $J_2(x,\hat u)$ is continuous on $\Gamma \times \SU$ and $\SU$ is compact, it follows from \cite[Theorem 7]{hogan1973point} that $\beta$ is continuous on $\Gamma$.
Thus, \eqref{eq:Gamma_def} implies $\gamma$ is continuous on $\Gamma$.

For all $x \in \Gamma$, define $b(x) \triangleq [1-\sigma(\gamma(x))] u_\rmb(x) + \sigma(\gamma(x)) u_*(x)$.
Since $u_*$, $\gamma$, and $u_\rmb$ are continuous on $\Gamma$, and $\sigma$ is continuous on $\BBR$, it follows that $b$ is continuous on $\Gamma$.
Next, let $c \in \mbox{bd } \Gamma$. 
Since $u_*(c) \in \SU$ is bounded, it follows from~\eqref{eq:gamma_def} and~\eqref{eq:Gamma_def} that $b(c) = u_{\rmb}(c)$.
Since $b$ is continuous on $\Gamma$, $u_\rmb$ is continuous on $\BBR^n$, and for all $x \in \mbox{bd } \Gamma$, $b(x) = u_{\rmb}(x)$, it follows from~\eqref{eq:u_cases_softmin} that $u$ is continuous on $\BBR^n$.

To prove~\ref{thm:softmin_u_cond}, let $d \in \BBR^n$. 
Since $u_\rmb(d) , u_*(d) \in \SU$, it follows from \eqref{eq:SU} that $A_u u_\rmb(d) \preceq b_u$ and $A_u u_*(d) \preceq b_u$.
Since, in addition, $\sigma(\gamma(d)) \in [0,1]$, it follows that 
\begin{gather}
[1-\sigma(\gamma(d))] A_u u_\rmb(d) \preceq [1-\sigma(\gamma(d))] b_u,\label{eq:ub_bound_thm1}\\
\sigma(\gamma(x)) A_u u_*(d) \preceq \sigma(\gamma(d)) b_u.\label{eq:u*_bound_thm1}
\end{gather}
Next, summing~\eqref{eq:ub_bound_thm1} and~\eqref{eq:u*_bound_thm1} and using \eqref{eq:u_cases_softmin} yields $A_u u(d) \preceq b_u$, which implies $u(d) \in \SU$.

To prove~\ref{thm:softmin_return_to_S_star}, assume for contradiction that for all $\tau \in (0, T_\rms]$, $x(t_1+\tau) \not \in \bar \SSS_*$.
Since, in addition, $x(t_1) \in \rm{bd \,}\bar \SSS_*$, it follows from \eqref{eq:defS_star} that for all $\tau \in [0, T_\rms]$, $\bar h_*(x(t_1+\tau)) \le 0$. 
Thus, Proposition~\ref{fact:softmin_bound} implies for all $\tau \in [0, T_\rms]$, $h(x(t_1+\tau)) < \bar h_*(x(t_1+\tau)) \le 0$, which combined with~\eqref{eq:gamma_def} and~\eqref{eq:Gamma_def} implies $x(t_1+\tau) \not \in \Gamma$.
Next, \eqref{eq:u_cases_softmin} implies the for all $\tau \in [0, T_\rms]$, $u(x(t_1+\tau)) = u_\rmb(x(t_1+\tau))$.
Hence, Proposition~\ref{prop:fwd_invar_S*} implies $x(t_1+T_\rms) \in \bar \SSS_*$, which is a contradiction.

To prove~\ref{thm:softmin_forward_inv}, let $a \in \Gamma$, and~\eqref{eq:gamma_def} and~\eqref{eq:Gamma_def} impliy $h(a) \ge \epsilon \ge \frac{1}{2}T_\rms l_\phi l_\rms$.
Since, in addition, Proposition~\ref{fact:softmin_bound} implies $\bar h_*(a) > h(a)$, it follows that $\bar h_*(a) > \frac{1}{2}T_\rms l_\phi l_\rms$. 
Thus,~\eqref{eq:ubarS_*_def} implies $a \in \mbox{int } \ubar \SSS_*$, which implies $\Gamma \subset \ubar \SSS_* \subseteq \SSS_*$.

Let $t_3 \ge 0$, and assume for contradiction that $x(t_3) \notin \SSS_*$.
Since, in addition, $x_0 \in \SSS_*$ and $\Gamma \subset \SSS_*$, it follows that there exists $t_2 \in [0,t_3]$ such that  $x(t_2) \in \SSS_*$ and for all $\tau \in [t_2,t_3]$, $x(\tau) \not \in \Gamma$.
Thus, \eqref{eq:u_cases_softmin} implies for all $\tau \in [t_2,t_3]$, $u(x(\tau)) = u_{\rmb}(x(\tau))$. 
Since, in addition, $x(t_2) \in \SSS_{*}$, Proposition~\ref{prop:fwd_invar_Sc} implies $x(t_3) \in \SSS_*$, which is a contradiction.  
{\hfill$\Box$} 
\end{pf}
\vspace{-0.4em}

Parts~\ref{thm:softmin_u_continuity} and \ref{thm:softmin_u_cond} of Theorem~\ref{thm:softmin} guarantee that the control is continuous and admissible. 
Part~\ref{thm:softmin_forward_inv} states that if $\epsilon \ge \frac{1}{2} T_\rms l_\phi l_\rms$, then $\SSS_*$ is forward invariant under the control and $x$ is in $\SSS_\rms$ for all time. 
Part~\ref{thm:softmin_return_to_S_star} shows that for any choice of $\epsilon >0$, $x$ is in the safe set $\SSS_\rms$ at sample times $0,T_\rms,2T_\rms,3T_\rms,\ldots$.

The control~\eqref{eq:h_softmin_def}--\eqref{eq:u_cases_softmin} relies on the Lie derivatives in~\eqref{eq:feas_check} and~\eqref{eq:qp_softmin.b}.
To calculate $L_fh$ and $L_gh$, note that 
\begin{align}
\label{eq:softmin_grad_h}
h^\prime(x) &= \vphantom{\sum_{i=0}^N} \frac{h_\rmb^\prime(\phi(x,NT_\rms)) Q(x, NT_\rms)}{e^{-\rho_1(h(x) - h_\rmb(\phi(x,NT_\rms)))}} \nn\\
&\qquad  + \sum_{i=0}^N \frac{h_\rms^\prime(\phi(x,iT_\rms)) Q(x, iT_\rms)}{e^{-\rho_1 (h(x) -h_\rms(\phi(x,iT_\rms)))} },
\end{align}
where $Q : \BBR^n \times [0,\infty) \to \BBR^{n \times n}$ is defined by
$Q(x,\tau) \triangleq \frac{\partial \phi(x,\tau)}{\partial x}$.
Differentiating \eqref{eq:phi_def} with respect to $x$ yields
\begin{equation}\label{eq:Q_integral}
Q(x,\tau) = I + \int_0^\tau \tilde f^\prime (\phi(x,s)) Q (x,s) \, \rmd s.
\end{equation}
Next, differentiating \eqref{eq:Q_integral} with respect to $\tau$ yields 
\begin{equation}\label{eq:sensitivity_ode}
     \frac{\partial Q(x,\tau)}{\partial \tau} = \tilde f^\prime (\phi(x,\tau)) Q(x,\tau).
\end{equation}
Note that for all $x \in \BBR^n$, $Q(x, \tau)$ is the solution to \eqref{eq:sensitivity_ode}, where the initial condition is $Q(x, 0)= I$. 
Thus, for all $x \in \BBR^n$, $L_fh(x)$ and $L_gh(x)$ can be calculated from \eqref{eq:softmin_grad_h}, where $\phi(x,\tau)$ is the solution to \eqref{eq:dynamics} under $u_\rmb$ on the interval $\tau \in [0,T]$ with $\phi(x,0) = x$, and $Q(x, \tau)$ is the solution to \eqref{eq:sensitivity_ode} on the interval $\tau \in [0,T]$ with $Q(x, 0)= I$. 
In practice, these solutions can be computed numerically at the time instants where the control algorithm \eqref{eq:h_softmin_def}--\eqref{eq:u_cases_softmin} is executed (i.e., the time instants where the control is updated). 
Algorithm~\ref{alg:softmin} summarizes the implementation of~\eqref{eq:h_softmin_def}--\eqref{eq:u_cases_softmin}, where $\delta t > 0$ is the time increment for a zero-order-hold on the control. 

The control~\eqref{eq:h_softmin_def}--\eqref{eq:u_cases_softmin} involves the user-selected parameters $\rho_1,\alpha,\kappa_h,\kappa_\beta >0$.
Recall that large $\rho_1$ improves the soft-minimum approximation of the minimum but can also result in large $\| h^\prime(x) \|_2$, which can tend to cause $\| \dot u(x(t)) \|_2$ to be large.
The quadratic program \eqref{eq:qp_softmin} shows that small $\alpha$ results in more conservative behavior; specifically, $u$ deviates more from the desired control $u_\rmd$ in order to keep the state trajectory farther away from $\mbox{bd }\SSS$. 
The homotopy \eqref{eq:u_cases_softmin} and definition \eqref{eq:gamma_def} of $\gamma$ show that large $\kappa_h$ or $\kappa_\beta$ cause the control $u$ to deviate more from the optimal control $u_*$ to the backup control $u_\rmb$ if either the feasibility metric $\beta$ or the barrier function $h$ are small.

\begin{figure}[t!]
\center{\includegraphics[width=0.48\textwidth,clip=true,trim= 0.2in 0.33in 1.2in 1.0in] {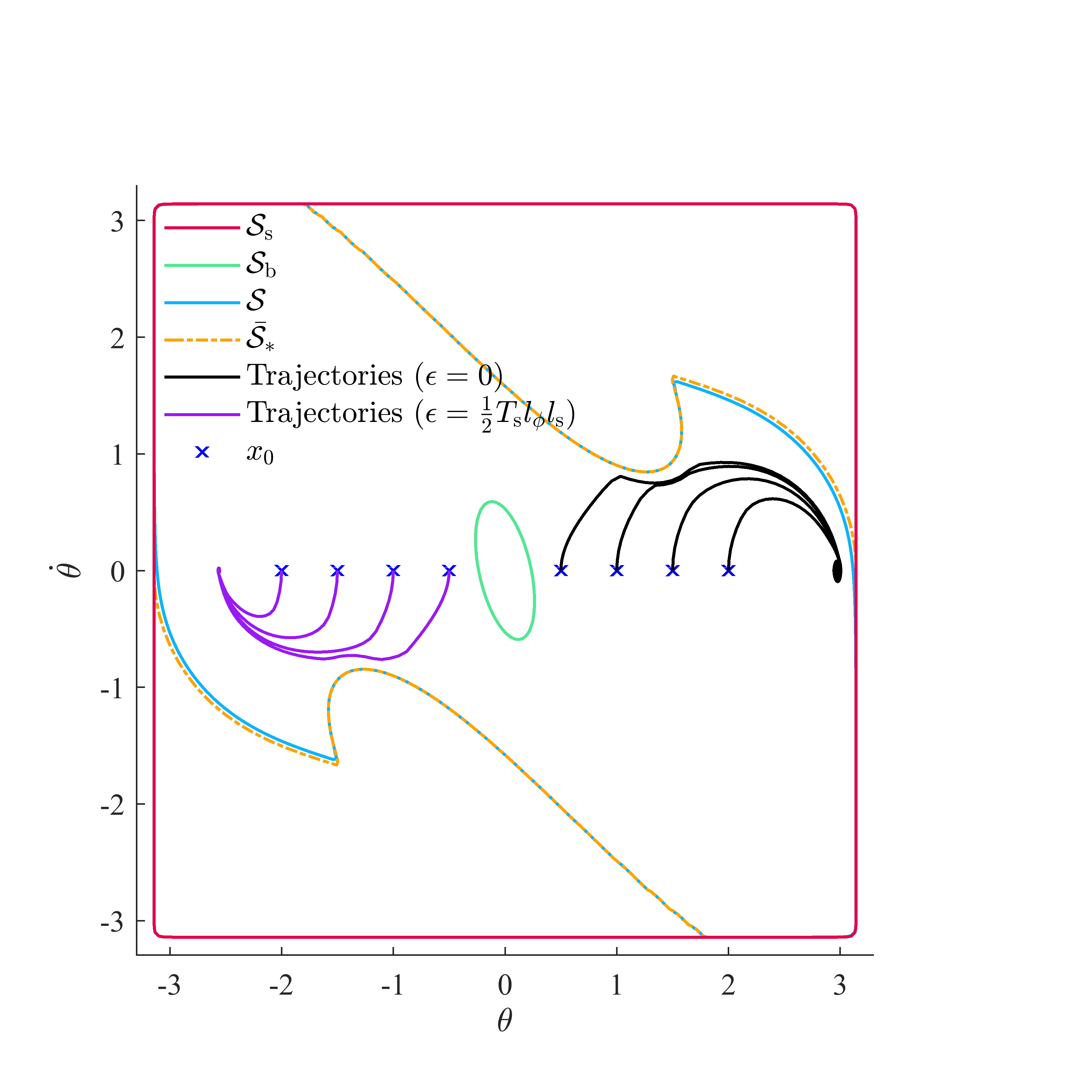}}
\caption{
$\SSS_\rms$, $\SSS_\rmb$, $\SSS$, $\bar \SSS_*$, and closed-loop trajectories for 8 initial conditions.}\label{fig:pendulum_traj}
\end{figure}

\begin{figure}[ht]
\center{\includegraphics[width=0.48\textwidth,clip=true,trim= 0.25in 0.25in 1.1in 0.6in] {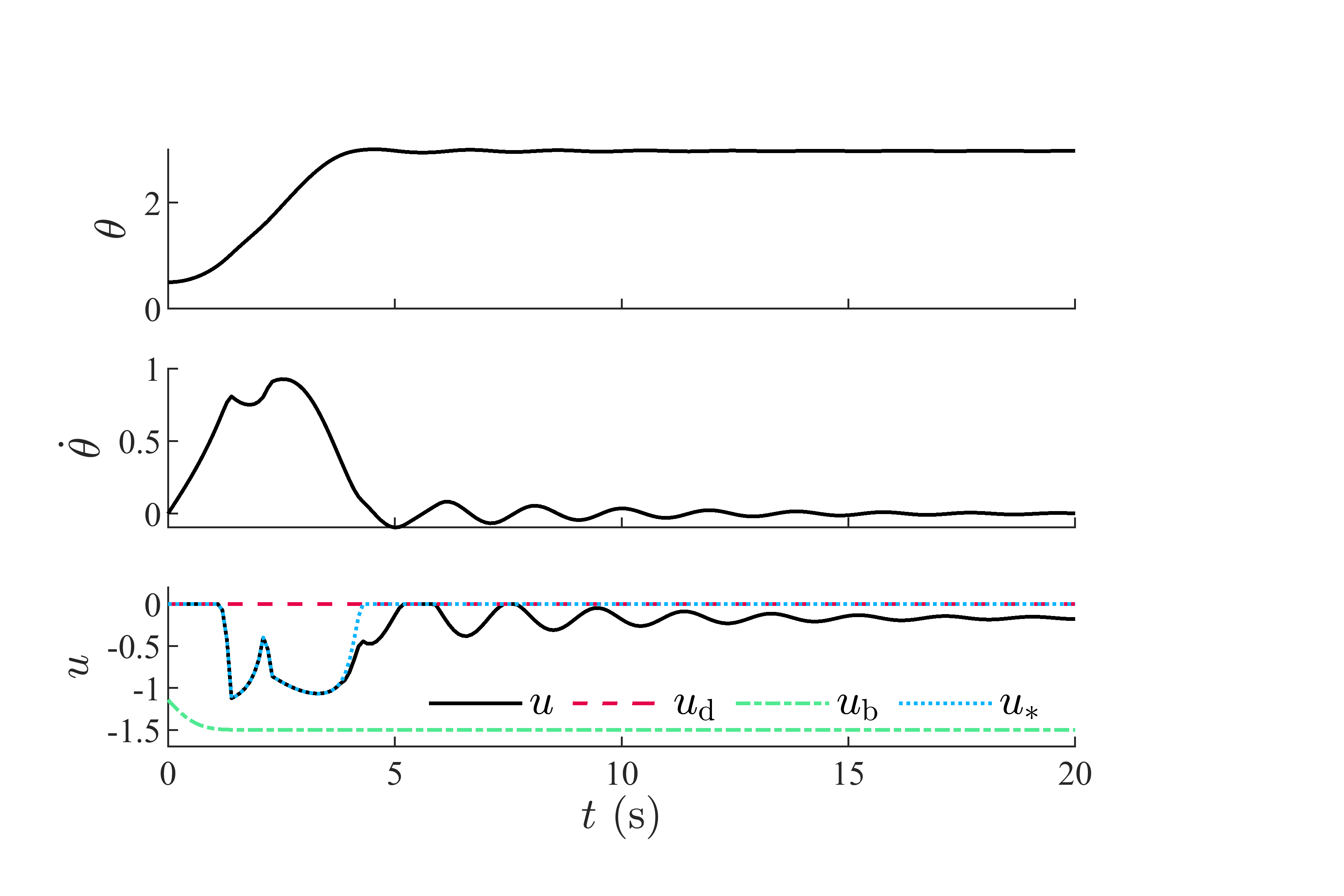}}
\caption{$\theta$, $\dot \theta$, $u$, $u_\rmd$, $u_\rmb$, and $u_*$ for $x_0=[0.5\,\,0]^\rmT$.}\label{fig:pendulum_states}
\end{figure}

\begin{figure}[ht]
\center{\includegraphics[width=0.48\textwidth,clip=true,trim= 0.15in 0.2in 1.1in 0.5in] {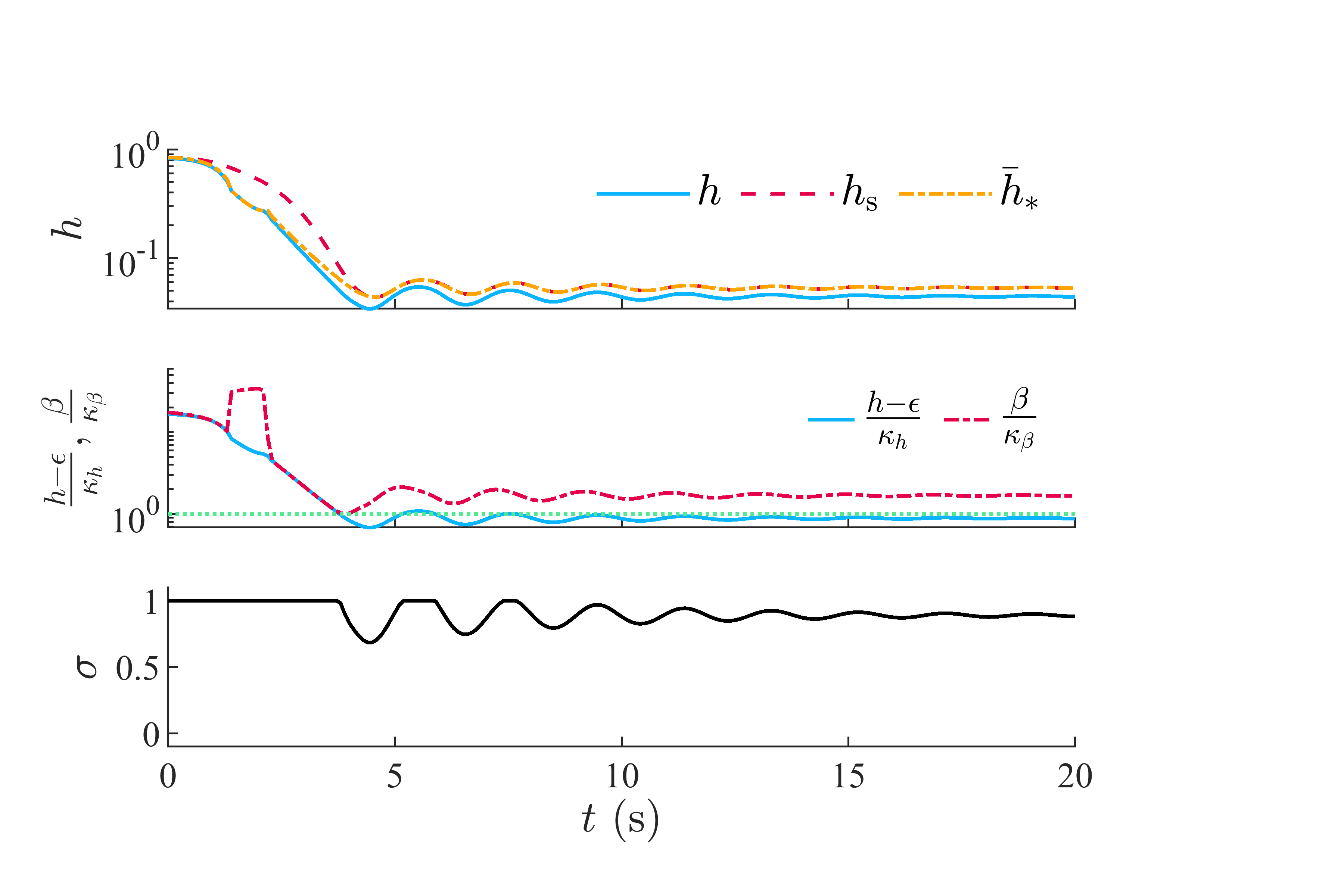}}
\caption{$h$, $h_\rms$, $\bar h_*$, $\frac{h-\epsilon}{\kappa_h}$, $\frac{\beta}{\kappa_\beta}$, and $\sigma$ for $x_0=[0.5\,\,0]^\rmT$.}\label{fig:pendulum_extra}
\end{figure}

\begin{algorithm}[ht]\label{alg:softmin}
\DontPrintSemicolon
\caption{Control using the soft-minimum BF quadratic program}
\KwIn{ $u_\rmd$, $u_\rmb$, $h_\rmb$, $h_\rms$, $N$, $T_\rms$, $\rho_1$, $\alpha$, $\epsilon$, $\kappa_h$, $\kappa_\beta$, $\sigma$, $\delta t$
}

\For{$k=0,1,2,\ldots$}{
    $x \gets x(k\delta t)$\;
    Solve~\eqref{eq:phi_def},~\eqref{eq:sensitivity_ode} for $\{\phi(x,iT_\rms)\}_{i=0}^N$, $\{Q(x,iT_\rms)\}_{i=0}^N$\;
    Compute $L_fh(x)$ and $L_gh(x)$ using~\eqref{eq:softmin_grad_h}\;
    $h \gets$ \eqref{eq:h_softmin_def},
    $\beta \gets$ \eqref{eq:feas_check},
    $\gamma \gets \min \{\frac{h- \epsilon}{\kappa_h}, \frac{\beta}{\kappa_\beta}\}$\;
    \eIf{$\gamma < 0$}{
        $u \gets u_\rmb(x)$\;}
    {
    $u_* \gets$ solution to quadratic program~\eqref{eq:qp_softmin}\;
    $u \gets [1-\sigma(\gamma)] u_\rmb(x) + \sigma(\gamma) u_*$\;
    }
}
\end{algorithm}
\begin{example}\label{ex:inverted_pendulum}
\rm 
Consider the inverted pendulum modeled by \eqref{eq:dynamics}, where 
\begin{equation*}
    f(x) = \begin{bmatrix}
    \dot \theta \\
    \sin \theta
    \end{bmatrix}, 
    \qquad
    g(x) = \begin{bmatrix}
    0 \\
    1
    \end{bmatrix}, 
    \qquad 
    x = \begin{bmatrix}
    \theta \\
    \dot \theta
    \end{bmatrix}, 
\end{equation*}
and $\theta$ is the angle from the inverted equilibrium.
Let $\bar u = 1.5$ and $\SU = \{ u \in \BBR \colon u \in [-\bar u, \bar u] \}$.
The safe set $\SSS_\rms$ is given by \eqref{eq:S_s}, where $h_\rms(x) = \pi - \|x\|_p$, $\|\cdot\|_p$ is the $p$-norm, and $p = 100$. 
The backup control is $u_\rmb(x) = \tanh K x$, where $K = [ \, -3 \quad -3 \,]$.
The backup safe set $\SSS_{\rmb}$ is given by \eqref{eq:S_b}, where $h_\rmb(x) = 0.07 - x^\rmT \matls 1.25 &  0.25 \\ 0.25 & 0.25 \matrs x.$
Lyapunov's direct method can be used to confirm that Assumption~\ref{assump:S_b_contract} is satisfied. 
The desired control is $u_\rmd = 0$, which implies that the objective is to stay in $\SSS_\rms$ using instantaneously minimum control effort. 
We implement the control~\eqref{eq:h_softmin_def}--\eqref{eq:u_cases_softmin} using $\rho_1 = 100$, $\alpha =1$, $\kappa_h = \kappa_\beta = 0.05$, and $\sigma$ given by Example~\ref{ex:sigma}. 
We let $\delta t= 0.1$~s, $N = 50$ and $T_\rms = 0.1$~s, which implies that the time horizon is $T = 5$~s.

Figure~\ref{fig:pendulum_traj} shows $\SSS_\rms$, $\SSS_\rmb$, $\SSS$, and $\bar \SSS_*$. 
Note that $\SSS \subset \bar \SSS_*$.
Figure~\ref{fig:pendulum_traj} also provides the closed-loop trajectories for 8 initial conditions, specifically, $x_0 = [ \, \theta_0 \quad 0 \,]^\rmT$, where $\theta_0 \in \{ \pm 0.5, \pm 1, \pm 1.5, \pm 2\}$.
We let $\epsilon =0$ for the initial conditions with $\theta_0 \in \{  0.5, 1, 1.5, 2\}$, and we let $\epsilon = \frac{1}{2} T_\rms l_\phi l_\rms$ for $\theta_0 \in \{  -0.5, -1, -1.5, -2\}$, which are the reflection of the first 4 across the origin.
For the cases with $\epsilon = \frac{1}{2} T_\rms l_\phi l_\rms$, Theorem~\ref{thm:softmin} implies that $\SSS_*$ is forward invariant under the control~\eqref{eq:u_cases_softmin}.
The trajectories with $\epsilon = \frac{1}{2} T_\rms l_\phi l_\rms$ are more conservative than those with $\epsilon = 0$.

Figure~\ref{fig:pendulum_states} and~\ref{fig:pendulum_extra} provide time histories for the case where $x_0=[\, 0.5\quad 0 \, ]^\rmT$ and $\epsilon = 0$.
Figure~\ref{fig:pendulum_states} shows $\theta$, $\dot \theta$, $u$, $u_\rmd$, $u_\rmb$, and $u_*$. 
The first row of Figure~\ref{fig:pendulum_extra} shows that $h$, $h_\rms$, and $\bar h_*$ are nonnegative for all time. 
The second row of Figure~\ref{fig:pendulum_extra} shows $\frac{h-\epsilon}{\kappa_h}$ and $\frac{\beta}{\kappa_\beta}$. 
Note that $\beta$ is positive for all time, which implies that~\eqref{eq:qp_softmin} is feasible at all points along the closed-loop trajectory. 
Since $\gamma$ is positive for all time but is less than $1$ in steady state, it follows from  \eqref{eq:u_cases_softmin} that $u$ in steady state is a blend of $u_\rmb$ and $u_*$. 
\exampletriangle
\end{example}

\begin{figure}[ht]
\center{\includegraphics[width=0.45\textwidth,clip=true,trim= 0.2in 0.23in 1.2in 1.0in] {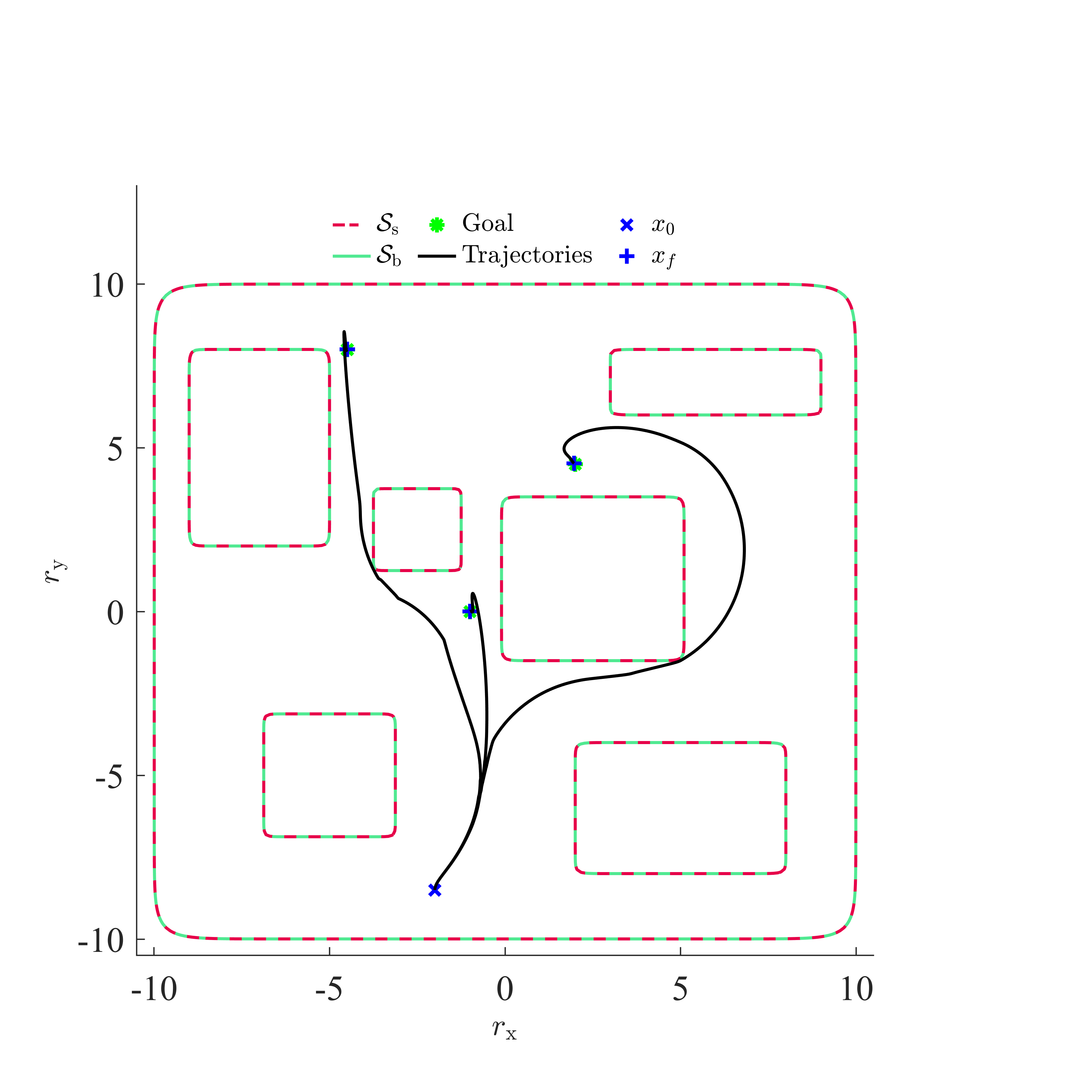}}
\caption{$\SSS_\rms$, $\SSS_\rmb$, and 3 closed-loop trajectories.}\label{fig:unicycle_new_map}
\end{figure}

\begin{figure}[ht]
\center{\includegraphics[width=0.45\textwidth,clip=true,trim= 0.3in 0.35in 1.1in 0.7in] {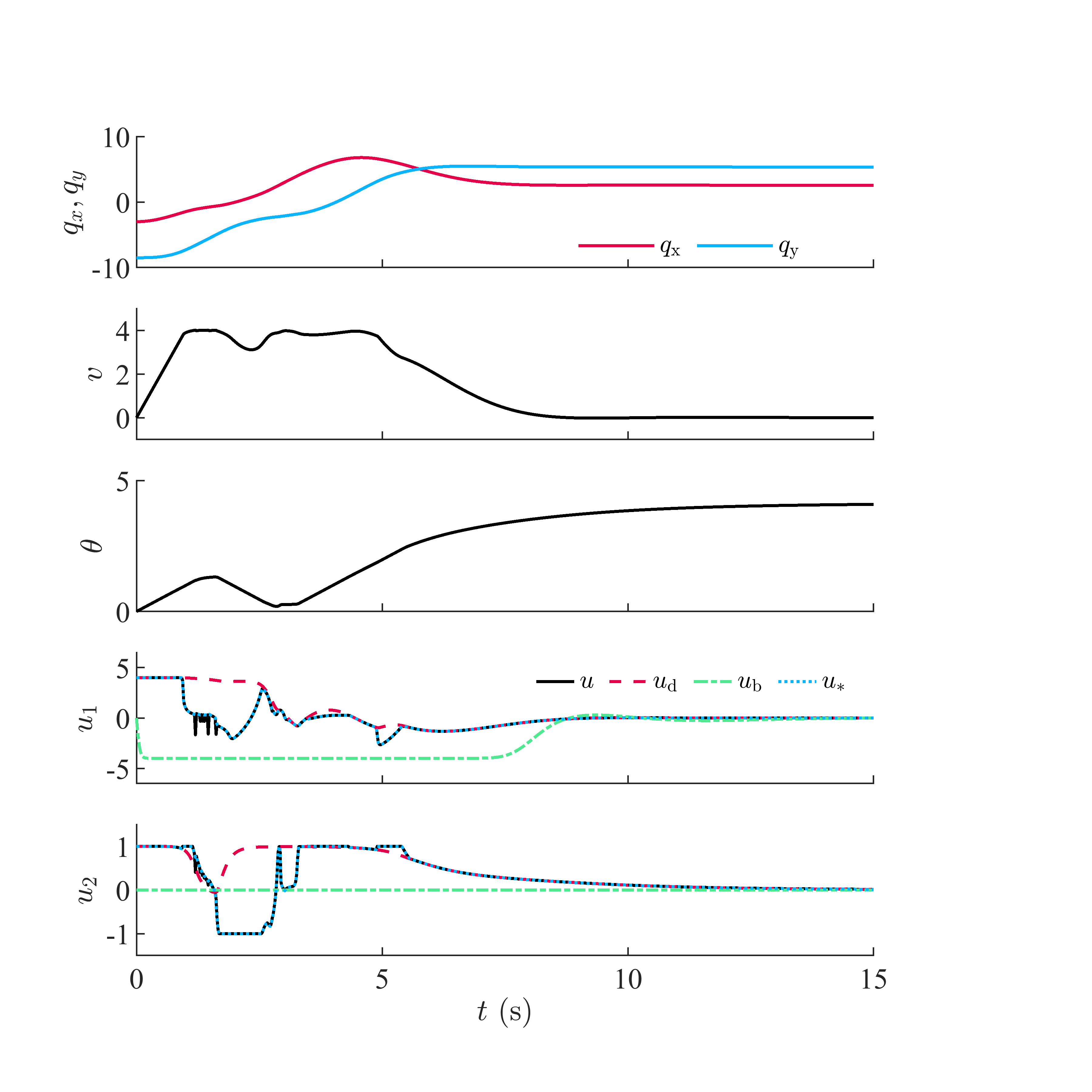}}
\caption{$q_\rmx$, $q_\rmy$, $v$, $\theta$, $u$, $u_\rmd$, $u_\rmb$, and $u_*$ for $r_{\rmd}=[\,2 \quad 4.5\,]^\rmT$.}\label{fig:unicycle_new_states_softmin_2}
\end{figure}

\begin{figure}[ht]
\center{\includegraphics[width=0.45\textwidth,clip=true,trim= 0.15in 0.25in 1.1in 0.55in] {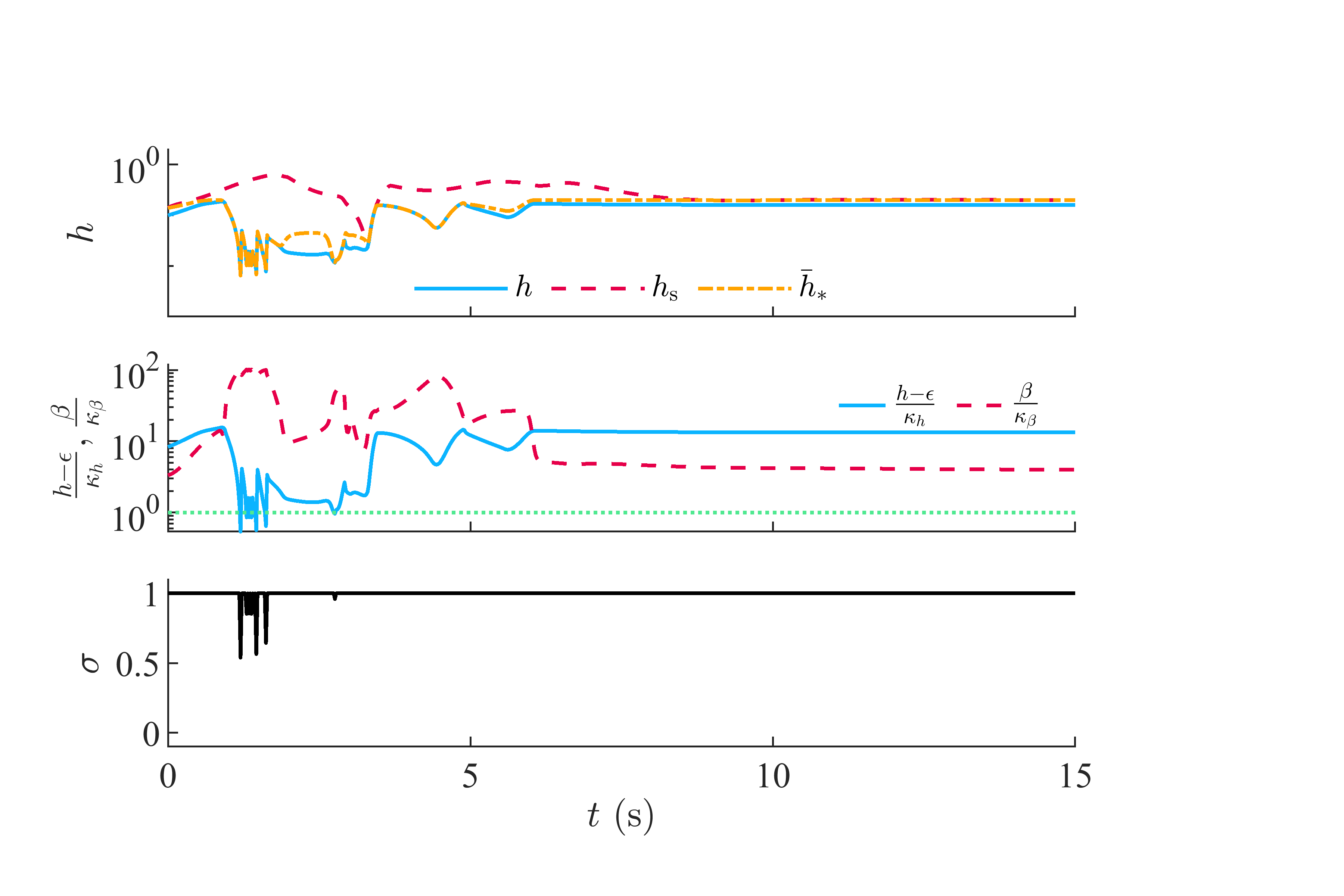}}
\caption{$h$, $h_\rms$, $\bar h_*$, $\frac{h-\epsilon}{\kappa_h}$, $\frac{\beta}{\kappa_\beta}$, and $\sigma$ for $r_{\rmd}=[\,2 \quad 4.5\,]^\rmT$.} \label{fig:unicycle_new_extra_softmin_2}
\end{figure}

\begin{example}\label{ex:ground_robot}\rm
Consider the nonholonomic ground robot modeled by~\eqref{eq:dynamics}, where
\begin{equation*}
    f(x) = \begin{bmatrix}
    v \cos{\theta} \\
    v \sin{\theta} \\
    0 \\
    0
    \end{bmatrix}, 
    \,
    g(x) = \begin{bmatrix}
    0 & 0\\
    0 & 0\\
    1 & 0 \\
    0 & 1
    \end{bmatrix}, 
    \,
    x = \begin{bmatrix}
    q_\rmx\\
    q_\rmy\\
    v\\
    \theta
    \end{bmatrix}, 
    \,
    u = \begin{bmatrix}
    u_1\\
    u_2
    \end{bmatrix}, 
\end{equation*}
and $[ \, q_\rmx \quad q_\rmy \, ]^\rmT$ is the robot's position in an orthogonal coordinate system, $v$ is the speed, and $\theta$ is direction of the velocity vector (i.e., the angle from $[ \, 1 \quad 0 \, ]^\rmT$ to $[ \, \dot q_\rmx \quad \dot q_\rmy \, ]^\rmT$). 
Let $\bar u_1 = 4$, $\bar u_2 = 1$, and 
$$\SU = \{ [ u_1 \, u_2 ]^\rmT \in\BBR^2: u_1 \in [-\bar u_1, \bar u_1], u_2 \in [-\bar u_2, \bar u_2] \}.$$
Define $r_\rmx \triangleq q_\rmx + d \cos{\theta}$, and $r_\rmy \triangleq q_\rmy + d \sin{\theta}$,
where $d=1$, and note that $[ \, r_\rmx \quad r_\rmy \, ]^\rmT$
is the position of a point-of-interest on the robot.
Consider the map shown in Figure~\ref{fig:unicycle_new_map}, which has 6 obstacles and a wall. For $i\in \{1,\ldots,6\}$, the area outside the $i$th obstacle is modeled as the zero-superlevel set of 
\begin{equation*}
q_i(x) = \left \| \matls a_{\rmx, i} (r_\rmx - b_{\rmx,i})  \\ a_{\rmy, i} (r_\rmy - b_{\rmy, i}) \\ a_{v,i}(v - b_{v,i}) \matrs \right \|_p - c_i,    
\end{equation*}
where $b_{\rmx,i}, b_{\rmy,i}, b_{v,i}, a_{\rmx,i}, a_{\rmy,i}, a_{v,i}, c_i, p> 0$ specify the location and dimensions of the $i$th obstacle.
The area inside the wall is modeled as the zero-superlevel set of
$$q_\rmw(x) = c - \left \| \matls a_{\rmx} r_\rmx  \\ a_{\rmy}r_\rmy \\ a_{v}v \matrs \right \|_p,$$
where $a_{\rmx}, a_{\rmy}, a_{v},c , p> 0$ specify the dimension of the space inside the wall.
The safe set $\SSS_\rms$ is given by \eqref{eq:S_s}, where $h_\rms(x) = \mbox{softmin}_{20} (q_{\rm w}(x),q_1(x), \ldots, q_6(x))$.
The safe set $\SSS_\rms$ projected into the $r_\rmx$--$r_\rmy$ plane is shown in Figure~\ref{fig:unicycle_new_map}.
Note that $\SSS_\rms$ is also bounded in speed $v$, specifically, for all $x \in \SSS_\rms$, $v\in[-1,9]$.

The backup control is $u_{\rmb}(x) = [\,\bar u_1 \tanh \mu v\quad 0\,]^\rmT$, 
where $\mu = -15$.
The backup safe set $\SSS_{\rmb}$ is given by \eqref{eq:S_b}, where $h_{\rmb}(x) = h_\rms(x) - 100 \frac{v^2}{\bar u_1}$.
Lyapunov's direct method can be used to confirm that Assumption~\ref{assump:S_b_contract} is satisfied. 
Figure~\ref{fig:unicycle_new_map} shows the projection of $\SSS_{\rmb}$ into the $r_\rmx$--$r_\rmy$ plane.

Let $r_{\rmd} \in \BBR^2$ be the goal location, that is, the desired location for $[ \, r_\rmx \quad r_\rmy \, ]^\rmT$.
Next, the desired control is
$u_{\rmd}(x) = [\,\bar u_1 \tanh v_{\rmd} (x)\quad \bar u_2 \tanh \omega_{\rmd}(x)\,]^\rmT$,
where
\begin{align*}
v_{\rmd}(x) &\triangleq -(\mu_1+\mu_2) v - (1+ \mu_1 \mu_2) e_{1}(x) + \frac{\mu_1^2}{d} e_{2}(x)^2,\\
\omega_{\rmd}(x) &\triangleq -\frac{\mu_1}{d}e_{2}(x),\\
\begin{bmatrix}
e_1(x) \\
e_2(x)
\end{bmatrix}
&\triangleq
\begin{bmatrix}
      \cos{\theta} & \sin{\theta}\\
    -\sin{\theta} & \cos{\theta}
\end{bmatrix}
\left (
\begin{bmatrix}
r_\rmx\\
r_\rmy 
\end{bmatrix} - r_\rmd \right ),
\end{align*}
where $\mu_1 = \mu_2 = 0.8$.
Note that the desired control is designed using a process similar to \cite[pp. 30--31]{de2002control}.

We implement the control~\eqref{eq:h_softmin_def}--\eqref{eq:u_cases_softmin} using $\rho_1 = 50$, $\alpha =1$, $\epsilon=0$, $\kappa_h = 0.012$, $\kappa_\beta =0.05$, and $\sigma$ given by Example~\ref{ex:sigma}.
We let $\delta t= 0.02\,\rms$, $N = 50$ and $T_\rms = 0.02\,\rms$.

Figure~\ref{fig:unicycle_new_map} shows the closed-loop trajectories for $x_0=[\,-3\quad -8.5\quad 0\quad 0\,]^\rmT$ with 3 different goal locations $r_{\rmd}=[\,2\quad 4.5\,]^\rmT$, $r_{\rmd}=[\,-1\quad 0\,]^\rmT$, and $r_{\rmd}=[\,-4.5\quad 8\,]^\rmT$. 
 In all cases, the robot position converges to the goal location while satisfying safety and the actuator constraints.

Figures~\ref{fig:unicycle_new_states_softmin_2} and~\ref{fig:unicycle_new_extra_softmin_2} show the trajectories of the relevant signals for the case where $r_{\rmd}=[\,2 \quad 4.5\,]^\rmT$.
Figure~\ref{fig:unicycle_new_extra_softmin_2}
shows that $\beta$ is positive for all time, which implies that~\eqref{eq:qp_softmin} is feasible at all points along the closed-loop trajectory. 
Since $\gamma$ is positive for all time and is greater than $1$ for $t > 3.4~\rms$, it follows from \eqref{eq:u_cases_softmin} that $u$ in steady state is equal to $u_*$ (as shown in Figure~\ref{fig:pendulum_states}). 
\exampletriangle
\end{example}

\section{Soft-Maximum/Soft-Minimum Barrier Function with Multiple Backup Controls}\label{section:softmax_softmin}

This section extends the method from the previous section to allow for multiple backup controls by adopting a soft-maximum/soft-minimum BF.
The following example illustrates the limitations of a single backup control and motivates the potential benefit of considering multiple backup controls. 

\begin{figure}[ht]
\center{\includegraphics[width=0.48\textwidth,clip=true,trim= 0.1in 1.25in 1.0in 2.0in] {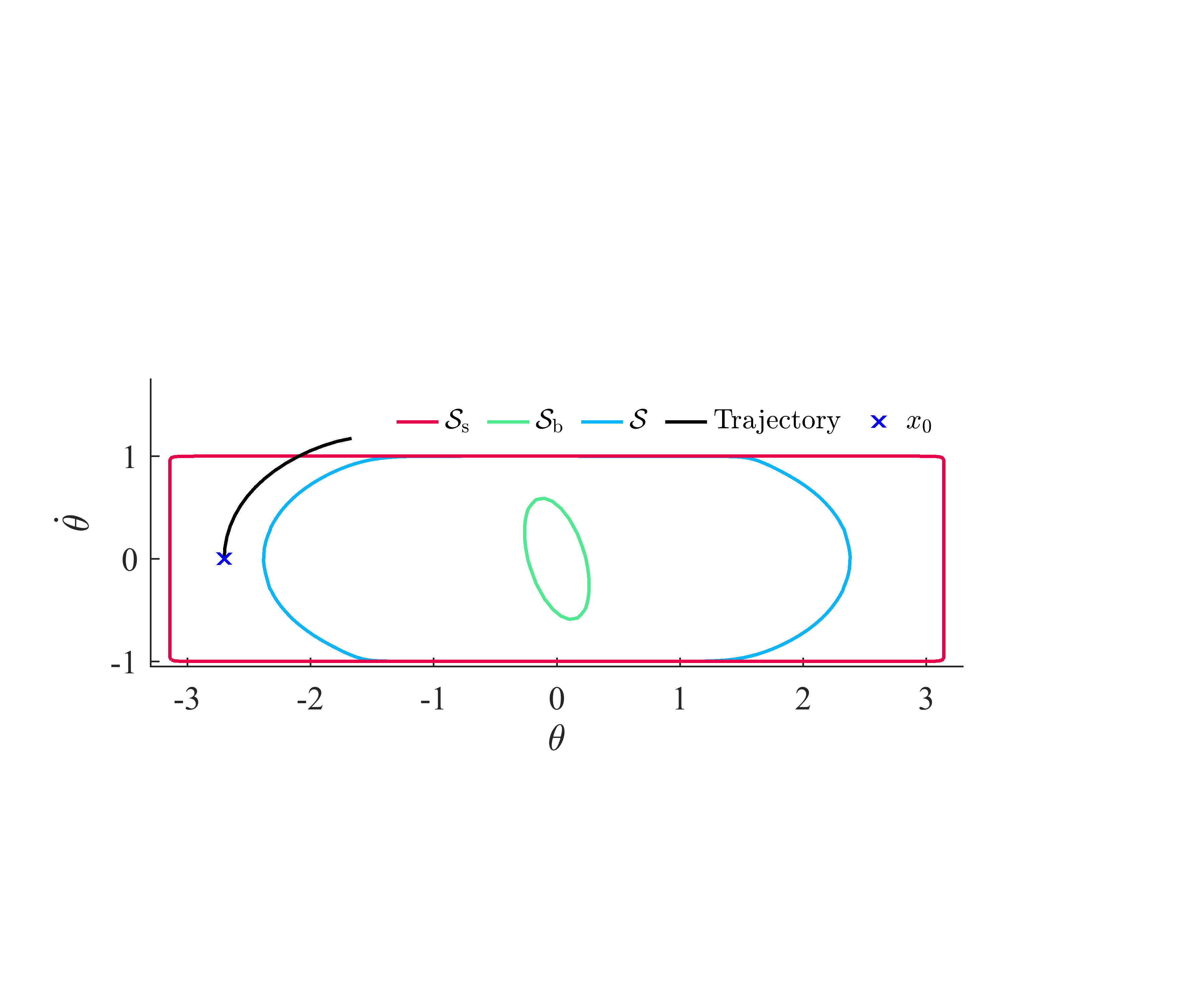}}
\caption{
$\SSS_\rms$, $\SSS_\rmb$, $\SSS$, $\bar \SSS_*$, and the closed-loop trajectory under Algorithm~\ref{alg:softmin} with $x_0 = [\,-2.7\,\,0\,]^\rmT$.} \label{fig:pendulum_motiv}
\end{figure}

\begin{example} \label{ex:inverted_pendulum_softmin_small}\rm
We revisit the inverted pendulum from Example~\ref{ex:inverted_pendulum}, where the safe set $\SSS_\rms$ is given by \eqref{eq:S_s}, where
$h_\rms(x) = 1 - \left \| \matls \frac{1}{\pi} & 0 \\ 0 & 1 \matrs x \right \|_{100}$.
We let $\epsilon=0$, $N=150$ and $T_\rms = 0.1$. 
Everything else is the same as in Example~\ref{ex:inverted_pendulum}. 

Figure~\ref{fig:pendulum_motiv} shows $\SSS_\rms$, $\SSS_\rmb$ and $\SSS$. 
We note that increasing $T$ does not change $\SSS$. 
In other words, $T$ was selected to yield the largest possible $\SSS$ under the backup control and safe set considered. 
Thus, with only one backup control, $\SSS$ cannot always be expanded by increasing $T$.

Figure~\ref{fig:pendulum_motiv} also shows the closed-loop trajectory under Algorithm~\ref{alg:softmin} with $x_0 = [ \, -2.7 \quad 0 \,]^\rmT$.
The state leaves the safe set $\SSS_\rms$.
In this example, $u = u_\rmb$ because that state is never in $\SSS$. 
\exampletriangle
\end{example}

This section presents a method to expand $\SSS$ by using multiple backup controls. 
Let $\nu$ be a positive integer, and consider the continuously differentiable backup controls $u_{\rmb_1}, \ldots, u_{\rmb_\nu}: \BBR^n \to \SU$. 
Let $h_{\rmb_j} : \BBR^n \to \BBR$ be continuously differentiable, and define the backup safe set
\begin{equation}
\label{eq:S_b_j}
\SSS_{\rmb_j} \triangleq \{x \in \BBR^n \colon h_{\rmb_j}(x) \geq 0 \}.  
\end{equation}
We assume $\SSS_{\rmb_j} \subseteq \SSS_\rms$ and make the following assumption.

\begin{assumption}\label{assump:S_b_contract_multi}\rm
For all $j \in \{1,\ldots,\nu\}$, if $u=u_{\rmb_j}$ and $x_0 \in \SSS_{\rmb_j}$, then for all $t \ge 0$, $x(t) \in \SSS_{\rmb_j}$. 
\end{assumption}

Assumption~\ref{assump:S_b_contract_multi} states that $\SSS_{\rmb_j}$ is forward invariant with respect to \eqref{eq:dynamics} where $u = u_{\rmb_j}$.
This is equivalent to Assumption~\ref{assump:S_b_contract} for each backup control.

Let $\tilde f_j:\BBR^n \to \BBR^n$ be defined by \eqref{eq:ftilde}, where $\tilde f$ and $u_\rmb$ are replaced by $\tilde f_j$ and $u_{\rmb_j}$.
Similarly, let $\phi_j:\BBR^n \times [0,\infty) \to \BBR^n$ be defined by \eqref{eq:phi_def}, where $\phi$ and $\tilde f$ are replaced by $\phi_j$ and $\tilde f_j$. 
Thus, $\phi_j(x,\tau)$ is the solution to \eqref{eq:dynamics} at time $\tau$ with $u=u_{\rmb_j}$ and initial condition $x$.

Let $T >0$, and consider $h_{*_j}, h_*:\BBR^n \to \BBR$ defined by
\begin{align*}
h_{*_j}(x) &\triangleq \min \, \left \{  h_{\rmb_j}(\phi_j(x,T)), \min_{\tau \in [0,T]} h_\rms(\phi_j(x,\tau))\right \}, \\
h_*(x) &\triangleq \max_{j \in \{1,\ldots, \nu\} } \, h_{*_j}(x), 
\end{align*}
and define $\SSS_{*_j} \triangleq \{ x \in \BBR^n \colon h_{*_j}(x) \ge 0 \}$ and $\SSS_* \triangleq \{ x \in \BBR^n \colon h_*(x) \ge 0 \}$.
The next result examines forward invariance of $\SSS_*$ and is a consequence of Proposition~\ref{prop:fwd_invar_Sc}.

\begin{proposition}\label{prop:fwd_invar_Sc.2}
\rm 
Let $\ell\in\{1,\ldots,\nu\}$, and consider \eqref{eq:dynamics}, where $x_0 \in \SSS_{*_\ell}$ and $u = u_{\rmb_\ell}$ satisfies Assumption~\ref{assump:S_b_contract_multi}.
Then, the following hold:
\begin{enumalph}

\item For all $t \ge T$, $x(t) \in \SSS_{\rmb_\ell}$. 

\item For all $t \ge 0$, $x(t) \in \SSS_{*_\ell} \subseteq \SSS_*$. 

\end{enumalph}
\end{proposition}

Next, let $N$ be a positive integer, and define $\SN \triangleq \{ 0, 1, \ldots, N \}$ and $T_\rms \triangleq T/N$.  
Then, consider $\bar h_{*_j}, \bar h_*:\BBR^n \to \BBR$ defined by
\begin{align}
\bar h_{*_j}(x) &\triangleq \min \Big \{ h_{\rmb_j}(\phi_j(x,NT_\rms)),  \min_{i \in \SN} h_\rms(\phi_j(x,iT_\rms)) \Big \},\label{eq:h_min_j_def} \\
\label{eq:h_max_min_def}
\bar h_*(x) &\triangleq \max_{j \in \{1,\ldots, \nu\} } \, \bar h_{*_j}(x),
\end{align}
and define
\begin{align}
\label{eq:defS_star_j}
\bar \SSS_{*_j} & \triangleq \{ x \in \BBR^n \colon \bar h_{*_j}(x) \ge 0 \},\\
\label{eq:defS_star_max}
\bar \SSS_* & \triangleq \{ x \in \BBR^n \colon \bar h_*(x) \ge 0 \}.
\end{align}
The next result is a consequence of Proposition~\ref{prop:fwd_invar_S*}.

\begin{proposition} \label{prop:fwd_invar_S*.2}
\rm 
Let $\ell\in\{1,\ldots,\nu\}$, and consider \eqref{eq:dynamics}, where $x_0 \in \SSS_{*_\ell}$ and $u = u_{\rmb_\ell}$ satisfies Assumption~\ref{assump:S_b_contract_multi}.
Then, the following hold:
\begin{enumalph}

\item For all $t \ge T$, $x(t) \in \SSS_{\rmb_\ell}$. 

\item For all $i\in \{ 0,1,\ldots,N \}$, $x(iT_\rms) \in \bar \SSS_{*_\ell} \subseteq \bar \SSS_*$. 

\end{enumalph}
\end{proposition}

Part \textit{(b)} of Proposition~\ref{prop:fwd_invar_S*.2} does not provide information regarding the state in between sample times.
Thus, we adopt an approach similar to that in Section~\ref{section:softmin}. 
Specifically, define the superlevel sets
\begin{align}
\label{eq:ubarS_*j_def_softmax}
\ubar \SSS_{*_j} &\triangleq \left \{x\in \BBR^n : \bar h_{*_j}(x) \ge \tfrac{1}{2} T_\rms l_\phi l_\rms \right \},\\
\label{eq:ubarS_*_def_softmax}
\ubar \SSS_* &\triangleq \left \{x\in \BBR^n : \bar h_*(x) \ge \tfrac{1}{2} T_\rms l_\phi l_\rms \right \},
\end{align}
where $l_\rms$ is the Lipschitz constant of $h_\rms$ with respect to the two norm and $l_\phi \triangleq \max_{j \in \{1,\ldots,\nu\}}\sup_{x\in \bar \SSS_{*_j}} \| \tilde f_j(x) \|_2$.
The next result is analogous to Proposition~\ref{prop:set_rel_softmin}, and its proof is similar.

\begin{proposition}\label{prop:set_rel_softmax}
\rm
The following statements hold:
\begin{enumalph}

\item For $j\in\{1,2,\ldots,\nu\}$, $\ubar \SSS_{*_j} \subseteq \SSS_{*_j} \subseteq \bar \SSS_{*_j} \subseteq\SSS_\rms$.

\item $\ubar \SSS_* \subseteq \SSS_* \subseteq \bar \SSS_* \subseteq\SSS_\rms$.

\end{enumalph}
\end{proposition}

Next, we use the soft minimum and soft maximum to define continuously differentiable approximations to $\bar h_{*_j}$ and $\bar h_*$.
Let $\rho_1,\rho_2 > 0$, and consider $h_j, h:\BBR^n \to \BBR$ defined by
\begin{align}
    h_j(x) &\triangleq \mbox{softmin}_{\rho_1} ( h_\rms(\phi_j(x,0)), h_\rms(\phi_j(x,T_\rms)), \ldots,\nn\\ 
    &\qquad h_\rms(\phi_j(x,NT_\rms)), h_\rmb(\phi_j(x,NT_\rms))),
    \label{eq:sofmin_h_j_def}
    \\
    h(x) &\triangleq \mbox{softmax}_{\rho_2} (h_1(x), \ldots, h_\nu(x)).\label{eq:h_softmaxmin_def}
\end{align}
and define
\begin{equation}\label{eq:defS_softmaxmin}
    \SSS \triangleq \{ x \in \BBR^n \colon h(x) \ge 0 \}.
\end{equation}
Proposition~\ref{fact:softmin_bound} implies that $\SSS \subset \bar \SSS_*$.

Let $\alpha >0$, $\epsilon\in [0, \max_{x\in\SSS} h(x))$, and $\kappa_h, \kappa_\beta > 0$.
Furthermore, let $\beta$, $\SB$, $\gamma$, $\Gamma$, and $u_*$ be given by~\eqref{eq:feas_check}--\eqref{eq:qp_softmin} where $h$ is given by~\eqref{eq:h_softmaxmin_def} instead of~\eqref{eq:h_softmin_def}.

We cannot use the control \eqref{eq:u_cases_softmin} because there are $\nu$ different backup controls rather than just one. 
Next, define
\begin{equation}
    \SSS_\epsilon  \triangleq \{ x \in \BBR^n : \bar h_*(x) \ge \epsilon \}, \label{eq:S_eps_def}
\end{equation}
and for all $x \in \bar \SSS_*$, define
\begin{equation}\label{eq:def_I}
I(x) \triangleq \{j: \bar h_{*_j}(x) \ge \epsilon \}.
\end{equation}
Then, for all $x \in \mbox{int } \SSS_\epsilon$, define the augmented backup control
\begin{equation}\label{eq:def_ua}
u_\rma(x) \triangleq \dfrac{\sum_{j \in I(x)} [\bar h_{*_j} (x) - \epsilon] u_{\rmb_j}(x)}{\sum_{j \in I(x)} [\bar h_{*_j}(x) - \epsilon]},
\end{equation}
which is a weighted sum of the backup controls for which $\bar h_{*_j}(x) > \epsilon$. 
Note that \eqref{eq:S_eps_def} implies that for all $x \in \mbox{int }  \SSS_\epsilon$, $I(x)$ is not empty and thus, $u_\rma(x)$ is well-defined.

\begin{proposition}\label{prop:u_a_cont}\rm
$u_\rma$ is continuous on $\mbox{int } \SSS_\epsilon$.
\end{proposition}
\vspace{-0.7em}
\begin{pf}
It follows from~\eqref{eq:def_I} and~\eqref{eq:def_ua} that $u_\rma(x) = n_\rma(x)/d_\rma(x)$, where 
\begin{align*}\label{eq:u_a_cont_proof}
n_\rma(x)  &\triangleq \sum_{j \in \{1,\ldots,\nu\}} \max\,\{0,\bar h_{*_j} (x) - \epsilon\} u_{\rmb_j}(x),\\
d_\rma(x)  &\triangleq \sum_{j \in \{1,\ldots,\nu\}} \max\,\{0, \bar h_{*_j}(x) - \epsilon\}.
\end{align*}
Since $u_{\rmb_j}$ and $\bar h_{*_j}$ are continuous on $\BBR^n$, it follows that $n_\rma$ and $d_\rma$ are continuous on $\BBR^n$. 
Since, in addition, for all $x \in \mbox{int } \SSS_\epsilon$, $d_\rma(x) \ne 0$, it follows that $u_\rma$ is continuous on $\mbox{int }\SSS_\epsilon$.
{\hfill$\Box$}
\end{pf}

The next result relates $\Gamma$ to $\mbox{int } \SSS_\epsilon$ and is a consequence of Proposition~\ref{fact:softmin_bound}, \eqref{eq:gamma_def}, \eqref{eq:Gamma_def}, and \eqref{eq:S_eps_def}. 

\begin{proposition}\label{prop:Gamma_subset}\rm
$\Gamma \subseteq \mbox{int } \SSS_\epsilon$.
\end{proposition}

Next, for all $x \in \Gamma \subseteq \mbox{int }\SSS_\epsilon$, define
\begin{equation}\label{eq:def_um}
u_\rmm(x) \triangleq[1-\sigma(\gamma(x))] u_\rma(x) + \sigma(\gamma(x)) u_*(x),
\end{equation}
which is the same as the homotopy in \eqref{eq:u_cases_softmin} except that $u_\rmb$ is replaced by the augmented backup control $u_\rma$.

Finally, consider the control 
\begin{equation}\label{eq:u_softmax_softmin}
u(x) = \begin{cases} 
u_\rmm(x),& \mbox{if } x \in \Gamma, \\ 
u_\rma(x), & \mbox{if } x \in \mbox{int } \SSS_\epsilon \backslash \Gamma,\\
u_{\rmb_q}(x),& \mbox{else},
\end{cases}
\end{equation}
where $q:[0,\infty) \to \{1,2,\ldots,\nu\}$ satisfies 
\begin{equation}\label{eq:def_q}
\begin{cases} 
\dot q = 0, & \mbox{if } x \not \in \mbox{bd } \SSS_\epsilon,\\
q^+ \in I(x), & \mbox{if } x \in \mbox{bd } \SSS_\epsilon, \\
\end{cases}
\end{equation}
where $q(0) \in \{1,\ldots,\nu\}$ and $q^+$ is the value of $q$ after an instantaneous change. 
It follows from \eqref{eq:def_q} that if $x \not \in \SSS_\epsilon$, then the index $q$ is constant.
In this case, the same backup control $u_{\rmb_q}$ is used in 
\eqref{eq:u_softmax_softmin} until the state reaches $\mbox{bd } \SSS_\epsilon$.
This approach is adopted so that switching between backup controls (i.e., switching $q$ in \eqref{eq:def_q}) only occurs on $\mbox{bd } \SSS_\epsilon$.

If there is only one backup control (i.e., $\nu=1$), then $u_\rma = u_\rmb$ and $u_{\rmb_q} = u_\rmb$.
In this case, the control in this section simplifies to the control~\eqref{eq:h_softmin_def}--\eqref{eq:u_cases_softmin} in Section~\ref{section:softmin}.

The following theorem is the main result on the soft-maximum/soft-minimum BF approach.

\begin{theorem}\label{thm:softmax_softmin}
\rm 
Consider \eqref{eq:dynamics} and $u$ given by~\eqref{eq:feas_check}--\eqref{eq:qp_softmin} and~\eqref{eq:sofmin_h_j_def}--\eqref{eq:def_q}, where $\SU$ given by \eqref{eq:SU} is bounded and nonempty, and $u_{\rmb_1}, \ldots, u_{\rmb_\nu}$ satisfy Assumption~\ref{assump:S_b_contract_multi}.
Then, the following hold:
\begin{enumalph}

\item 
$u$ is continuous on $\BBR^n \backslash \mbox{bd } \SSS_\epsilon$.
\label{thm:softmax_u_continuity}

\item 
For all $x \in \BBR^n$, $u(x) \in \SU$. 
\label{thm:softmax_u_cond}

\item 
\label{thm:softmax_return_to_S_star_new}
Let $x_0 \in \bar \SSS_*$ and $q(0) \in \{ j \colon \bar h_{*_j}(x_0) \ge 0 \}$.
Assume there exists $t_1 \ge 0$ such that $x(t_1) \in \mbox{bd } \bar \SSS_*$. 
Then, there exists $\tau \in (0, T_\rms]$ such that $x(t_1+\tau) \in \bar \SSS_*\subseteq \SSS_\rms$.

\item 
\label{thm:softmax_forward_inv}
Let $\epsilon \ge \frac{1}{2} T_\rms l_\phi l_\rms$, $x_0 \in \SSS_*$, and $q(0) \in \{ j \colon h_{*_j}(x_0) \ge 0 \}$.
Then, for all $t \ge 0$, $x(t) \in \SSS_* \subseteq \SSS_\rms$.

\end{enumalph}
\end{theorem}

\begin{pf}
To prove~\ref{thm:softmax_u_continuity}, note that the same arguments in the proof to Theorem~\ref{thm:softmin}\ref{thm:softmin_u_continuity} imply that $u_*$ and $\gamma$ are continuous on $\Gamma$.
Next, Propositions~\ref{prop:u_a_cont} and~\ref{prop:Gamma_subset} imply that $u_\rma$ is continuous on $\Gamma$. 
Since, in addition, $\sigma$ is continuous on $\BBR$, it follows from \eqref{eq:def_um} that $u_\rmm$ is continuous on $\Gamma$.
Next, let $c \in \mbox{bd } \Gamma$. 
Since $u_*(c) \in \SU$ is bounded, it follows from~\eqref{eq:gamma_def}, \eqref{eq:Gamma_def}, and \eqref{eq:def_um} that $u_\rmm(c) = u_{\rma}(c)$.
Since $u_\rmm$ is continuous on $\Gamma$, $u_\rma$ is continuous on $\mbox{int }\SSS_\epsilon$, and for all $x \in \mbox{bd } \Gamma$, $u_\rmm(x) = u_{\rma}(x)$, it follows from~\eqref{eq:u_cases_softmin} that $u$ is continuous on $\mbox{int } \SSS_\epsilon$.
Next, \eqref{eq:def_q} implies for $x \in \BBR^n \backslash \SSS_\epsilon$, $q$ is constant.
Since, in addition, $u_{\rmb_j}$ is continuous on $\BBR^n$, it follows from \eqref{eq:u_softmax_softmin} that $u$ is continuous on $\BBR^n \backslash \SSS_\epsilon$.
Thus, $u$ is continuous on $(\mbox{int } \SSS_\epsilon) \cup ( \BBR^n \backslash \SSS_\epsilon)$, which confirms~\ref{thm:softmax_u_continuity}.

To prove~\ref{thm:softmax_u_cond}, let $d\in \BBR^n$, and we consider 2 cases: $d\in \BBR^n \backslash \mbox{int } \SSS_\epsilon$, and $d \in \mbox{int } \SSS_\epsilon$. First, let $d\in \BBR^n \backslash \mbox{int } \SSS_\epsilon$, and \eqref{eq:u_softmax_softmin} implies $u(d) = u_{\rmb_q}(d)\in\SU$.
Next, let $d \in \mbox{int } \SSS_\epsilon$.
Since for $j\in\{1,\ldots,\nu\}$, $u_{\rmb_j}(d) \in\SU$, it follows from~\eqref{eq:def_ua} that $u_\rma(d) \in \SU$. 
Since $u_\rma(d),u_*(d) \in \SU$, the same arguments in the proof to Theorem~\ref{thm:softmin}\ref{thm:softmin_u_cond} with $u_\rmb$ replaced by $u_\rma$ imply that $u(d) \in \SU$, which confirms~\ref{thm:softmax_u_cond}.

To prove~\ref{thm:softmax_return_to_S_star_new}, assume for contradiction that for all $\tau \in (0, T_\rms]$, $x(t_1+\tau) \not \in \bar \SSS_*$, and it follows from \eqref{eq:defS_star_max} and \eqref{eq:S_eps_def} that for all $\tau \in (0, T_\rms]$, $x(t_1+\tau) \not \in \SSS_\epsilon$.
Next, we consider two cases: (i) there exists $t \in [0,t_1]$ such that $x(t) \in \mbox{bd } \SSS_\epsilon$, and (ii) for all $t \in [0,t_1]$, $x(t) \not \in \mbox{bd } \SSS_\epsilon$. 

First, consider case (i), and it follows that there exists $t_\rmi \in [0,t_1]$ such that $x(t_\rmi) \in \mbox{bd } \SSS_\epsilon$ and for all $\tau \in (t_\rmi, t_1+T_\rms]$, $x(\tau) \not \in \SSS_\epsilon$. 
Hence, \eqref{eq:def_q} and \eqref{eq:u_softmax_softmin} imply that there exists $\ell \in I(x(t_\rmi))$ such that for all $\tau \in [t_\rmi,t_1+T_\rms]$, $q(\tau) = \ell$ and $u(x(\tau)) = u_{\rmb_\ell}(x(\tau))$. 
Next, let $N_\rmi$ be the positive integer such that $t_\rmi+N_\rmi T_\rms \in (t_1,t_1+T_\rms]$, and define $\tau_\rmi \triangleq t_\rmi+N_\rmi T_\rms - t_1 \in (0,T_\rms]$. 
Since $x(t_\rmi) \in \SSS_{\epsilon_\ell} \subseteq \bar \SSS_{*_\ell}$ and for all $\tau \in [t_\rmi,t_1+T_\rms]$, $u(x(\tau)) = u_{\rmb_\ell}(x(\tau))$, it follows from Proposition~\ref{prop:fwd_invar_S*.2} that $x(t_\rmi + N_\rmi T_\rms)=x(t_1+\tau_\rmi) \in \bar \SSS_*$, which is a contradiction.

Next, consider case (ii), and it follows that for all for all $\tau \in [0, t_1+T_\rms]$, $x(\tau) \not \in \SSS_\epsilon$. 
Hence, \eqref{eq:def_q} and \eqref{eq:u_softmax_softmin} imply that for all $\tau \in [0,t_1+T_\rms]$, $q(\tau) = q(0)$ and $u(x(\tau)) = u_{\rmb_{q(0)}}(x(\tau))$. 
Next, let $N_0$ be the positive integer such that $N_0 T_\rms \in (t_1,t_1+T_\rms]$, and define $\tau_0 \triangleq N_0 T_\rms - t_1 \in (0,T_\rms]$. 
Since $x_0 \in \SSS_{\epsilon_\ell} \subseteq \bar \SSS_{*_\ell}$ and for all $\tau \in [0,t_1+T_\rms]$, $u(x(\tau)) = u_{\rmb_\ell}(x(\tau))$, it follows from Proposition~\ref{prop:fwd_invar_S*.2} that $x(N_0 T_\rms)= x(t_1+\tau_0) \in \bar \SSS_*$, which is a contradiction.

To prove~\ref{thm:softmax_forward_inv}, since $\epsilon \ge \frac{1}{2}T_\rms l_\phi l_\rms$, it follows from~\eqref{eq:ubarS_*_def_softmax}, \eqref{eq:S_eps_def}, and Proposition~\ref{prop:set_rel_softmax} that $\SSS_\epsilon \subseteq \ubar \SSS_* \subseteq \SSS_*$. 
Define $\SSS_{\epsilon_j}  \triangleq \{ x \in \BBR^n : \bar h_{*_j}(x) \ge \epsilon \}$, and it follows from~\eqref{eq:ubarS_*j_def_softmax} and Proposition~\ref{prop:set_rel_softmax} that $\SSS_{\epsilon_j} \subseteq \ubar \SSS_{*_j} \subseteq \SSS_{*_j}$.

Let $t_3 \ge 0$, and assume for contradiction that $x(t_3) \not \in \SSS_*$, which implies  $x(t_3) \not \in \SSS_\epsilon$.
Next, we consider two cases: (i) there exists $t \in [0,t_3)$ such that $x(t) \in \SSS_\epsilon$, and (ii) for all $t \in [0,t_3)$, $x(t) \not \in \SSS_\epsilon$. 

First, consider case (i), and it follows that there exists $t_2 \in [0,t_3)$ such that $x(t_2) \in \mbox{bd } \SSS_\epsilon$ and for all $\tau \in (t_2,t_3]$, $x(\tau) \not \in \SSS_\epsilon$. 
Thus, \eqref{eq:def_q} and \eqref{eq:u_softmax_softmin} imply that there exists $\ell \in I(x(t_2))$ such that for all $\tau \in [t_2,t_3]$, $q(\tau) = \ell$ and  $u(x(\tau)) = u_{\rmb_\ell}(x(\tau))$. 
Since, in addition, $x(t_2) \in \SSS_{\epsilon_\ell} \subseteq \SSS_{*_\ell}$, it follows from Proposition~\ref{prop:fwd_invar_Sc.2} that $x(t_3) \in \SSS_*$, which is a contradiction. 

Next, consider case (ii), and \eqref{eq:u_softmax_softmin} and \eqref{eq:def_q} imply that for all $\tau \in [0,t_2]$, $q(\tau) = q(0)$ and $u(x(\tau)) = u_{\rmb_{q(0)}}(x(\tau))$. 
Since, in addition, $x_0 \in \SSS_{*_{q(0)}}$, Proposition~\ref{prop:fwd_invar_Sc.2} implies $x(t_2) \in \SSS_*$, which is a contradiction. 
{\hfill$\Box$}\end{pf}
\vspace{-1em}

Theorem~\ref{thm:softmax_softmin} provides the same results as Theorem~\ref{thm:softmin} except $u$ is not necessarily continuous on $\mbox{bd }\SSS_\epsilon$ because there are multiple backup controls. 
Specifically, $u$ is not continuous on $\{ x \in \mbox{bd }\SSS_\epsilon \colon I(x) \mbox{ is not a singleton} \}$; however, the following remark illustrates a condition under which $u$ is continuous at a point on $\mbox{bd }\SSS_\epsilon$.

\begin{remark}\rm
{Let $t_1 > 0$ such that $x(t_1) \in \mbox{bd } \SSS_\epsilon$, and let $t_1^-$ and $t_1^+$ denote times infinitesimally before and after $t_1$. 
If $x(t_1^-) \in \SSS_\epsilon$, $x(t_1^+) \notin \SSS_\epsilon$, and $I(x(t^-))$ is a singleton, then $u$ is continuous at $x(t_1)$.}
\end{remark}

The control~\eqref{eq:feas_check}--\eqref{eq:qp_softmin} and~\eqref{eq:sofmin_h_j_def}--\eqref{eq:def_q} can be computed using a process similar to the one described immediately before Example~\ref{ex:inverted_pendulum}.
Algorithm~\ref{alg:softmax_softmin} summarizes the implementation of~\eqref{eq:feas_check}--\eqref{eq:qp_softmin} and~\eqref{eq:sofmin_h_j_def}--\eqref{eq:def_q}.

\begin{algorithm}[t!]
\DontPrintSemicolon
\caption{Control using the soft-maximum and soft-minimum BF quadratic program with multiple backup controls}\label{alg:softmax_softmin}
\KwIn{ $u_\rmd$, $u_{\rmb_j}$, $h_{\rmb_j}$, $h_\rms$, $N$, $T_\rms$, $\rho_1$, $\rho_2$, $\alpha$, $\epsilon$, $\kappa_h$, $\kappa_\beta$, $\sigma$, $\delta t$ 
}
\For{$k=0,1,2\ldots$}{
    $x \gets x(k\delta t)$\;
    \For{j=$1,\ldots,\nu$}{
        $\{\phi_j(x,iT_\rms)\}_{i=0}^N, \{Q_j(x,iT_\rms)\}_{i=0}^N \gets$ \eqref{eq:phi_def}, \eqref{eq:sensitivity_ode}\;
        $\bar h_{*_j} \gets$ \eqref{eq:h_min_j_def},
        $h_j \gets$ \eqref{eq:sofmin_h_j_def}\;
        }
    $\bar h_* \gets$ \eqref{eq:h_max_min_def}\;
    \eIf{$\bar h_* \le \epsilon$}{
        $u \gets u_{\rmb_q}(x)$ where $q$ satisfies~\eqref{eq:def_q}
        }
    {
    Compute $L_fh(x)$ and $L_gh(x)$\;
    $h \gets$ \eqref{eq:h_softmaxmin_def}, 
    $\beta \gets$ \eqref{eq:feas_check}, 
    $\gamma \gets \min \{\frac{h - \epsilon}{\kappa_h}, \frac{\beta}{\kappa_\beta}\}$\;
    $u_\rma \gets$ \eqref{eq:def_ua}\;
    \eIf{$\gamma < 0$}{$u \gets u_{\rma}$}
    {
    $u_* \gets$ solution to quadratic program~\eqref{eq:qp_softmin}\;
    $u \gets [1-\sigma(\gamma)] u_\rma + \sigma(\gamma) u_*$\;}
    }
}
\end{algorithm}

\begin{figure}[t!]
\center{\includegraphics[width=0.45\textwidth,clip=true,trim= 0.15in 1.0in 1.2in 1.8in] {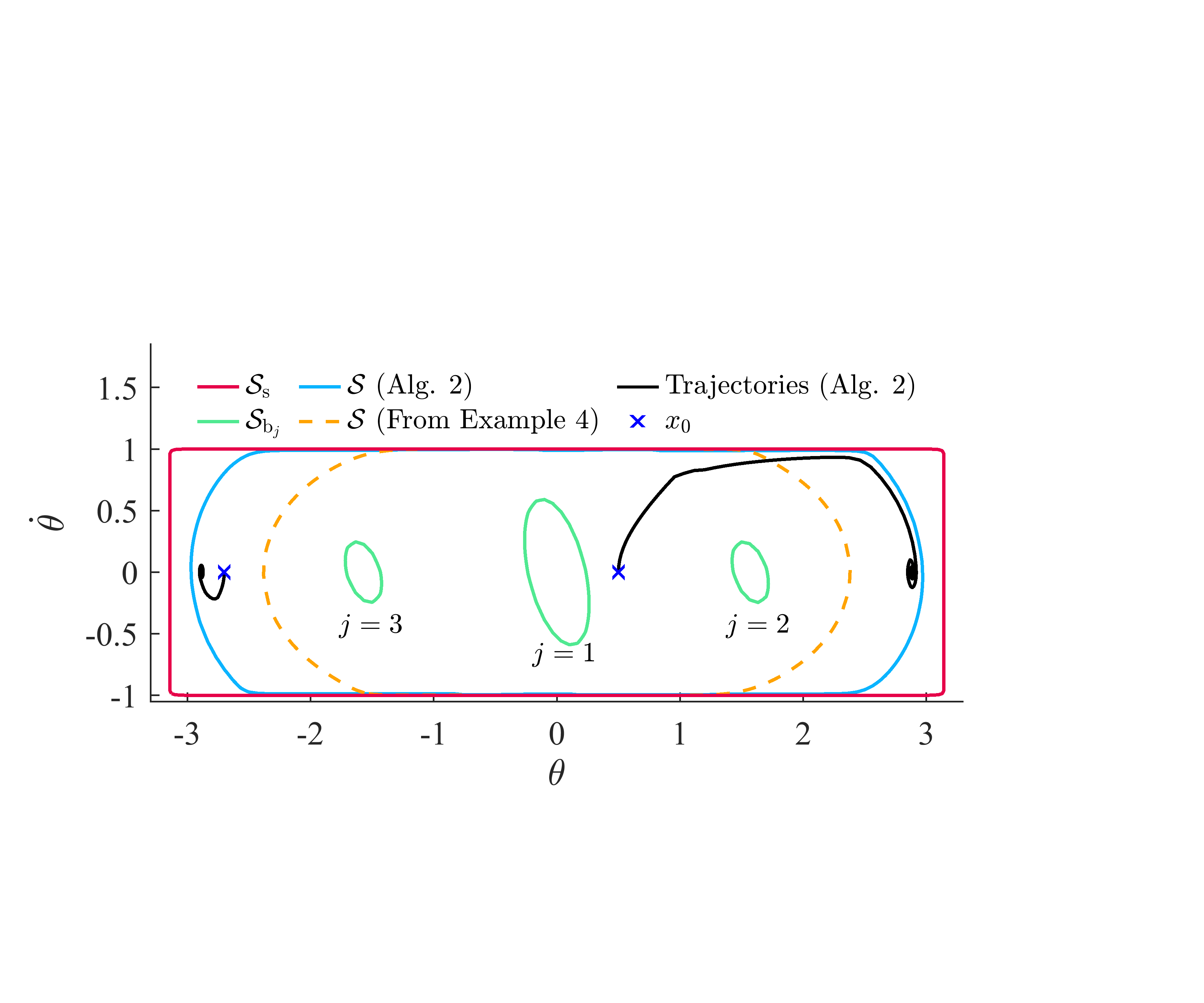}}
\caption{
$\SSS_\rms$, $\SSS_{\rmb_1}$, $\SSS_{\rmb_2}$, $\SSS_{\rmb_3}$, $\SSS$ with Algorithm~\ref{alg:softmax_softmin} , $\SSS$ from Example~\ref{ex:inverted_pendulum_softmin_small}, and closed-loop trajectories for 2 initial conditions.}\label{fig:pendulum_multi_traj}
\end{figure}

\begin{figure}[ht]
\center{\includegraphics[width=0.45\textwidth,clip=true,trim= 0.25in 0.25in 1.1in 0.6in] {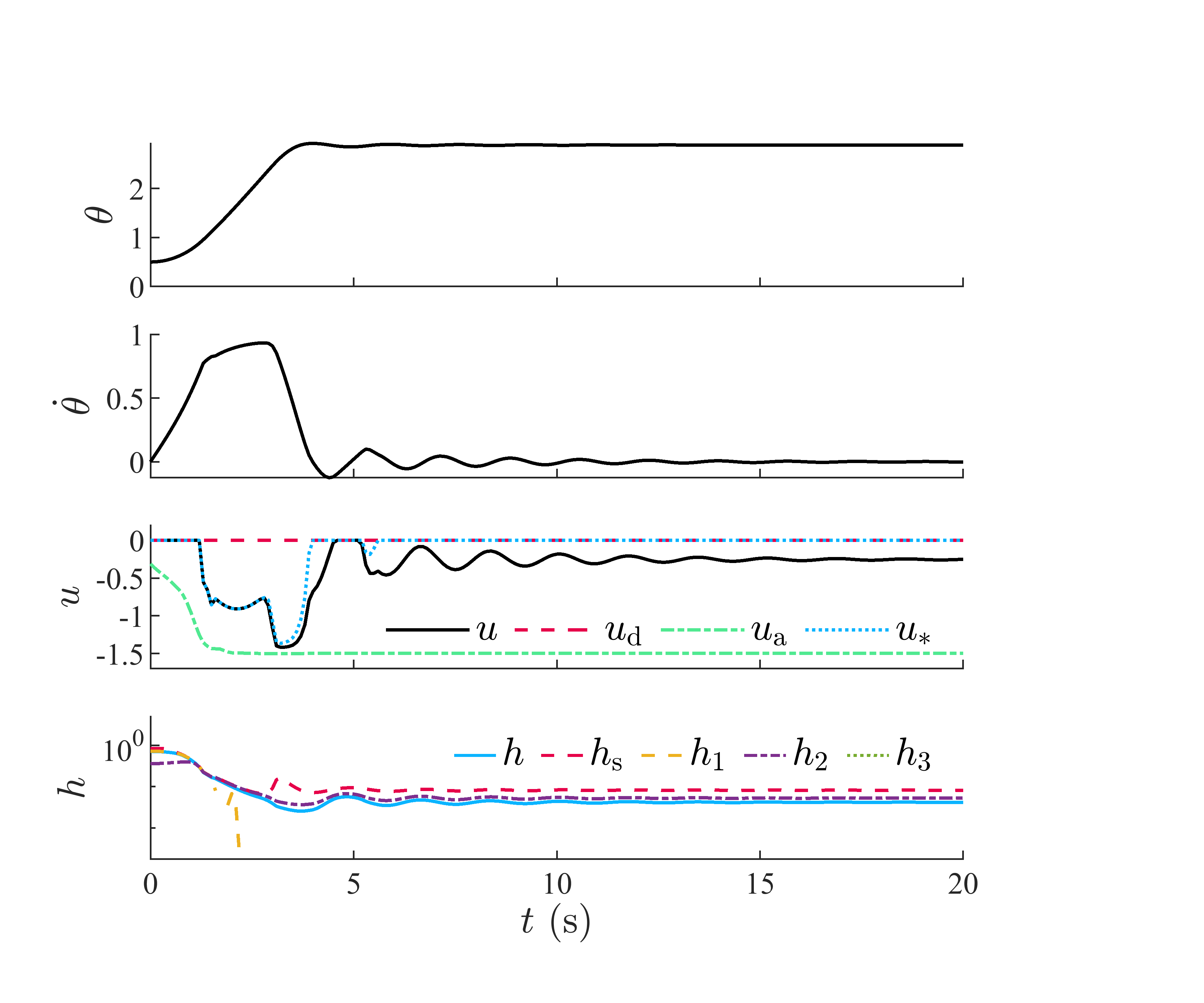}}
\caption{$\theta$, $\dot \theta$, $u$, $u_\rmd$, $u_\rma$, $u_*$, $h$, $h_\rms$, $h_1$, $h_2$, $h_3$ for $x_0=[0.5\,\,0]^\rmT$.}\label{fig:pendulum_multi_states}
\end{figure}

\begin{example}\label{ex:inverted_pendulum_softmax}\rm
We revisit the inverted pendulum from Example~\ref{ex:inverted_pendulum_softmin_small} but use multiple backup controls to enlarge $\SSS$ in comparison to Example~\ref{ex:inverted_pendulum_softmin_small}. 
The safe set $\SSS_\rms$ is the same as in Example~\ref{ex:inverted_pendulum_softmin_small}. 
For $j\in\{1,2, 3\}$, the backup controls are $u_{\rmb_j}(x) = \tanh K(x-x_{\rmb_j})$, where $x_{\rmb_1}\triangleq 0$, $x_{\rmb_2}\triangleq[\, {\pi}/{2}\quad 0 \, ]^\rmT$, $x_{\rmb_3}\triangleq[\, -{\pi}/{2}\quad 0 \, ]^\rmT$, and $K = [ \, -3 \quad -3 \,]$.
The backup safe sets are given by \eqref{eq:S_b_j}, where $$h_{\rmb_1}(x) = 0.07 - x^\rmT \matls 1.25 &  0.25 \\ 0.25 & 0.25 \matrs x,$$
and for $j\in\{2, 3\}$, 
$$h_{\rmb_j}(x) = 0.025 - (x - x_{\rmb_j})^\rmT \matls 1.17 &  0.17 \\ 0.12 & 0.22 \matrs (x - x_{\rmb_j}).$$
Note that $u_{\rmb_1}$ and $\SSS_{\rmb_1}$ are the backup control and backup safe set used in Example~\ref{ex:inverted_pendulum_softmin_small}. 
Lyapunov's direct method can be used to confirm that Assumption~\ref{assump:S_b_contract_multi} is satisfied. 
The desired control is $u_\rmd = 0$.

We implement the control~\eqref{eq:feas_check}--\eqref{eq:qp_softmin} and~\eqref{eq:sofmin_h_j_def}--\eqref{eq:def_q} using $\rho_2 = 50$ and the same parameters as in Example~\ref{ex:inverted_pendulum_softmin_small} except $N = 50$ rather than 150.
We selected $N = 50$ rather than 150 because this example has 3 backup controls, so $T$ was reduced by 1/3 to obtain a computational complexity that is comparable to Example~\ref{ex:inverted_pendulum_softmin_small},

Figure~\ref{fig:pendulum_multi_traj} shows $\SSS_\rms$, $\SSS_{\rmb_1}$, $\SSS_{\rmb_2}$, $\SSS_{\rmb_3}$, $\SSS$.
Note that $\SSS$ using multiple backup controls is larger than that $\SSS$ from Example~\ref{ex:inverted_pendulum_softmin_small}, which uses only one backup control and has a comparable computational cost.
Figure~\ref{fig:pendulum_multi_traj} also shows the closed-loop trajectories under Algorithm~\ref{alg:softmax_softmin} for 2 initial conditions, specifically, $x_0=[\,-2.7\quad 0]^\rmT$ and $x_0=[\,0.5\quad 0]^\rmT$.
Example~\ref{ex:inverted_pendulum_softmin_small} shows that the closed-loop trajectory leaves $\SSS_\rms$ under Algorithm~\ref{alg:softmin} with $x_0 = [ \, -2.7 \quad 0 \,]^\rmT$.
In contrast, Figure~\ref{fig:pendulum_multi_traj} shows that Algorithm~\ref{alg:softmax_softmin} keeps the state in $\SSS_\rms$. 

Figure~\ref{fig:pendulum_multi_states} provides time histories for the case where $x_0=[\,0.5\quad 0]^\rmT$.
The last row of Figure~\ref{fig:pendulum_multi_states} shows that $h$ and $h_\rms$ nonnegative for all time and that the soft maximum in $h$ is initially an approximation of $h_1$ and then becomes an approximation of $h_2$ as the trajectory moves closer to $\SSS_{\rmb_2}$.
Note that, $\gamma$ is positive for all time but is less than $1$ in steady state, it follows from \eqref{eq:u_softmax_softmin} that $u$ in steady state is a blend of $u_\rma$ and $u_*$.
\exampletriangle
\end{example}

\bibliographystyle{elsarticle-num}
\bibliography{backup_cbf}

\appendix
\end{document}